\documentclass[12pt]{article}
\usepackage[margin=25mm]{geometry}
\RequirePackage[OT1]{fontenc}
\RequirePackage{amsthm,amsmath}
\RequirePackage[colorlinks,citecolor=blue,urlcolor=blue]{hyperref}
\def\bm{\boldsymbol}
\usepackage{graphicx,psfrag,epsf,subcaption}
\usepackage{algorithm}
\usepackage{algpseudocode}%
\usepackage{enumerate}
\usepackage{enumitem}
\usepackage[sort&compress]{natbib}
\usepackage{amssymb}
\usepackage{bm}
\usepackage{url}
\usepackage{float}
\usepackage{siunitx}
\usepackage{bbm}
\usepackage{booktabs}
\usepackage{array}

\usepackage{comment}

\usepackage[toc,page]{appendix}
\usepackage[font=small,labelfont=bf]{caption}
\RequirePackage[normalem]{ulem} 

\graphicspath{{plots/}}
\let\hat\widehat

\ifx\BlackBox\undefined
\newcommand{\BlackBox}{\rule{1.5ex}{1.5ex}}  % end of proof
\fi

\ifx\QED\undefined
\def\QED{~\rule[-1pt]{5pt}{5pt}\par\medskip}
\fi

\ifx\proof\undefined
\newenvironment{proof}{\par\noindent{\bf Proof\ }}{\hfill\BlackBox\\[2mm]}
\fi

\ifx\theorem\undefined
\newtheorem{theorem}{Theorem}
\fi
\ifx\example\undefined
\newtheorem{example}[theorem]{Example}
\fi
\ifx\property\undefined
\newtheorem{property}{Property}
\fi
\ifx\lemma\undefined
\newtheorem{lemma}[theorem]{Lemma}
\fi
\ifx\proposition\undefined
\newtheorem{proposition}[theorem]{Proposition}
\fi
\ifx\remark\undefined
\newtheorem{remark}[theorem]{Remark}
\fi
\ifx\corollary\undefined
\newtheorem{corollary}[theorem]{Corollary}
\fi
\ifx\definition\undefined
\newtheorem{definition}[theorem]{Definition}
\fi
\ifx\conjecture\undefined
\newtheorem{conjecture}[theorem]{Conjecture}
\fi
\ifx\fact\undefined
\newtheorem{fact}[theorem]{Fact}
\fi
\ifx\claim\undefined
\newtheorem{claim}[theorem]{Claim}
\fi
\ifx\assumption\undefined

\fi
\numberwithin{equation}{section}
\numberwithin{theorem}{section}
\usepackage{multirow}

\newcommand{\E}{\mathbb{E}}

\newcommand{\mbf}[1]{\mathbf{#1}}
\newcommand{\sbf}[1]{\mathbf{#1}}

\renewcommand{\Pr}{\mathbb{P}}

\newcommand{\cX}{\mathcal{X}}

\newcommand{\cH}{\mathcal{H}}

\newcommand{\bE}{\mathbb{E}}

\newcommand{\cP}{\mathcal{P}}

\newcommand{\bbR}{\mathbb{R}}

\newcommand{\bs}{\bm{s}}

\newcommand{\bA}{\bm{A}}
\newcommand{\bB}{\bm{B}}
\newcommand{\bC}{\bm{C}}

\newcommand{\cM}{\mathcal{M}}

\newcommand{\II}{\mathbb{I}}

\newcommand{\PP}{\mathbb{P}}

\newcommand{\RR}{\mathbb{R}}

\def\bs{\boldsymbol}

\newcommand{\argmin}{\mathop{\mathrm{argmin}}}
\newcommand{\argmax}{\mathop{\mathrm{argmax}}}

\newcommand{\diag}{{\rm diag}}

\def\sang{\textcolor{blue}}

\newtheorem{assumption}{Assumption}

\newtheorem{lemmain}{Lemma}

\providecommand{\keywords}[1]
{
	\small	
	\textbf{\textit{Keywords---}} #1
}

\title{Variational Nonparametric Inference in Functional Stochastic Block Model}
\author{Zuofeng Shang$^{1}$, Peijun Sang$^{2}$, Yang Feng$^{3}$ and Chong Jin$^{1}$ \\
	\small $^{1}$ Department of Mathematical Sciences, New Jersey Institute of Technology\\
	\small $^{2}$Department of Statistics and Actuarial Science, University of Waterloo \\
		\small $^{3}$ School of Global Public Health, New York University
}
\begin{document}
\def\bigtimes{\mbox{\LARGE$\times$}}

\maketitle

\begin{abstract}
We propose a functional stochastic block model whose vertices involve functional data information. This new model extends the classic stochastic block model with vector-valued nodal information, and finds applications in real-world networks whose nodal information could be functional curves. Examples include international trade data in which a network vertex (country) is associated with the annual or quarterly GDP over certain time period, and MyFitnessPal data in which a network vertex (MyFitnessPal user) is associated with daily calorie information measured over certain time period. 
Two statistical tasks will be jointly executed. First, we will detect community structures of the network vertices assisted by the functional nodal information. Second, we propose computationally efficient variational test
to examine the significance of the functional nodal information. We show that the community detection algorithms achieve weak and strong consistency, and the variational test is asymptotically chi-square with diverging degrees of freedom. As a byproduct, we propose pointwise confidence intervals for the slop function of the functional nodal information. Our methods are examined through both simulated and real datasets.
\end{abstract}

\hspace{10pt}


\section{Introduction}

In classic graphical models where the nodes are represented by data vectors, namely, nodal information, a key assumption is that nodal dependence can be characterized by a sparse precision matrix.
Specifically, one draws an edge between nodes $i,j$ if the $(i,j)$-th entry of the precision matrix is nonzero.
This is naturally true in the Gaussian graphical model since the nodes $i,j$ are conditionally dependent, given the remaining nodal information, if and only if 
the $(i,j)$-entry of the precision matrix is nonzero.
Detection of dependence structure is naturally degenerated to the problem of estimating a sparse precision matrix, which is methodologically feasible by applying such as the renowned graphical Lasso approach (\citealp{yuan2007biometrika, friedman2007biostatistics, aspremont2008siam, rothman2008ejs, ravikumar2011ejs, yuan2010jmlr, cai2011jasa, liu2012aos, xue2012aos}) or certain nonconvex penalization approaches such as SCAD (\citealp{fan2009aos}). 
The above strategy has been recently extended to functional settings where the nodes are represented by random functions. In this scenario, detection of dependence structure is degenerated to the more challenging block precision matrix estimation through $\ell_1$-penalization (\citealp{qiao2019jasa, zapata2021biometrika}) or 
precision operator estimation through regularization (\citealp{li2018nonparametric, solea2022jasa}).

The relationship between nodal dependence and sparse precision matrix might be limited in practice.
A more direct way can characterize nodal dependence through
the stochastic block model (SBM), which assumes that the nodes are divided into (unknown) communities and the observed
edge variables between the nodes follow certain probability distributions. Thus, the SBM can neatly characterize both nodal dependence and community structure without resorting to the precision matrix.
In an SBM, the edge information is fully summarized in a so-called adjacency matrix; the community can be understood as a cluster of nodes that share some common features.
For example, in the classical SBM proposed by \cite{holland1983stochastic}, each node is assumed to belong to one community, and the $(i,j)$-th entry of the adjacency matrix is assumed to follow a Bernoulli distribution. Additionally, the probability of the presence of an edge between nodes $i,j$ solely depends on the community memberships of the two nodes, and nodes from the same community are more likely to be linked by an edge than nodes from different communities. A major objective when fitting an SBM is to identify the memberships of all nodes. Various estimation and inference methods have been proposed to fit this classical SBM; see \cite{bickel2013aos}, \cite{li2022hierarchical}, \cite{wang2023fast} and references therein. 

It should be noted that in the classical SBM, only the adjacency matrix has been taken into consideration to determine the community structure. Recently, \cite{weng2022} 
studied the more realistic scenario that each node is represented by a data vector. 
They found that leveraging the nodal information in a proper way can enhance the accuracy of community detection. In particular, in addition to the assumptions in the classical SBM, multivariate nodal information was assumed to affect the membership of each node through a multinomial regression, where the number of communities is a fixed priori. To alleviate the computational burden, they further proposed a variational produce that can execute the community detection task in a suitable time duration.

In light of the advantage of incorporating nodal information in an SBM, we consider a new SBM with functional nodal information, where longitudinal or functional observations are available at each node. The nodal information is related to the edge variables through an infinite-dimensional functional parameter, and thus, our model is nonparametric. In the international trading network studied in \cite{wang2023fast}, they only accounted for the amount of exports and imports between countries to determine the adjacency matrix, and then detected the membership of each country in the network based on the adjacency matrix. They identified three communities and found that the membership is consistent with the GDP level for most countries in this network. This finding motivates us to directly incorporate the GDP profile over more years of each country into a functional SBM model in order to improve community detection as well as explore the effect of GDP levels on the membership assignment of the countries through nonparametric testing. 
The functional SBM also finds applications in the MyFitnessPal dataset where the aim is to use daily calorie information to assist community detection as well as examine the significance of calorie information in membership assignment.  

Our contributions in this article are summarized as follows. Firstly, 
to the best of our knowledge, this is the first time that functional nodal information is accounted for statistical inference in an SBM, including community detection, functional parameter estimation and nonparametric testing. This is beneficial to future research on incorporating functional data analysis into network data analysis. 
Secondly, we develop an efficient algorithm based on the idea of variational inference to fit the proposed functional SBM. But it is worth noting that while the proposed variational method alleviates the computational burden when fitting the functional SBM, it leads to more challenges to investigate theoretical properties of the proposed methods. Our third contribution is to overcome these challenges and develop critical theoretical properties for our methods. In particular, we establish the convergence rate for the functional estimator; we develop a statistically sound variational testing procedure for the functional parameter and derive its asymptotic null distribution that leads to a valid testing rule. We further show that the estimated communities for network nodes achieves weak and strong consistency in the sense of \cite{bickel2009pnas} under mild conditions. Last but not least, we carry out extensive simulation and real-world studies to demonstrate that, compared with the classical SBM, accounting for functional covariates as nodal information in the proposed functional SBM can indeed enhance the accuracy of community detection. 

The remainder of the article is organized as follows. We introduce the functional SBM in Section \ref{sec:model} and the nonparametric testing problem in Section \ref{sec:testing}. We develop a variational estimation procedure to fit the model in Section \ref{sec:algarithms}. Theoretical properties of the proposed estimator are studied in Section \ref{sec:theory}. We perform extensive simulation studies to demonstrate the finite sample performance of the proposed method in Section \ref{sec:simulation}, and apply our method to the international trading network in Section \ref{sec:application}. Section \ref{sec:conclusion} concludes the article. All technical proofs and additional simulation results are relegated to the supplementary material.

\section{Functional Stochastic Block Model} \label{sec:model}

We consider an undirected network $(\mathcal{V},\mathcal{E})$ in which 
$\mathcal{V}=[n]:=\{1,2,\ldots,n\}$ is a set of $n$ nodes and
$\mathcal{E}$ is a set of edges.
The corresponding symmetric adjacency matrix $A\in\{0,1\}^{n\times n}$
satisfies $A_{ij}=1$ if and only if $(i,j)\in\mathcal{E}$.
Since $(i,i)\notin\mathcal{E}$ for any $i\in[n]$, this implies that $A_{ii}=0$, i.e.,
no self-loops are allowed. 
For $i\in[n]$, let $Z_i\in[K]:=\{0,1,\ldots,K\}$ denote the community assignment of node $i$ and $Z=(Z_1,\ldots,Z_n)$
be the assignment vector. Here, we have assumed that the $n$ nodes are partitioned into $K+1$ underlying communities in the network.
Let $X_i(t), t\in[0,1]$ be a functional covariate associated with node $i$
and $X=(X_1,\ldots,X_n)$.
For instance, in World GDP Data,
one can regard $X_i(t)$ as the GDP of country $i$ at time $t$, and $A_{ij}$ as the connectivity of countries $i$ and $j$;
in MyFitnessPal data, one can regard $X_i(t)$ as the difference
of the intake and planned calories for individual $i$ at time $t$, and $A_{ij}$ as the similarity of individuals $i$ and $j$.

Let $H^m[0,1]$ be the $m$th order Sobolev space on $[0,1]$ with $m\ge1$.
For $\alpha=(\alpha_1,\ldots,\alpha_K)^\top\in\bbR^K$ and $\beta=(\beta_1,\ldots,\beta_K)\in \cH_{m,K}:=(H^m[0,1])^K$,
$i\in[n]$, define
\begin{equation}\label{eqn:Wia}
W_{ia}(\alpha,\beta)=\left\{
\begin{array}{cc}\frac{e^{\alpha_a+\int_0^1X_i(t)\beta_{a}(t)dt}}{1+\sum_{k=1}^{K}
e^{\alpha_k+\int_0^1X_i(t)\beta_{k}(t)dt}},& 1\le a\le K,\\
\frac{1}{1+\sum_{k=1}^{\sang{K}}
e^{\alpha_k+\int_0^1X_i(t)\beta_{k}(t)dt}},& a=0.
\end{array}\right.
\end{equation}
We propose the following functional stochastic block model (fSBM):
\begin{eqnarray}
P(Z|X)&=&\prod_{i=1}^nW_{iZ_i}({\alpha}, \beta),\nonumber\\
P(A|Z)&=&\prod_{1\le i<j\le n}(B_{Z_iZ_j})^{A_{ij}}(1-B_{Z_iZ_j})^{1-A_{ij}},\label{model:eq1}
\end{eqnarray}
where $\bB=[B_{kk'}]_{k,k'=0}^{K}\in[0,1]^{(K+1)\times (K+1)}$ is an unknown block probability matrix,
$\alpha$
denotes the unknown vector of intercepts, and $\beta$ denotes the unknown vector of slope functions.
In particular, $\beta_k$ characterizes the functional relationship between $X_i(t)$ and $Z_i, A_{ij}$'s, given that node $i$ belongs to community $k$; a large magnitude of $\beta_k$ may indicate that even a small perturbation in $X_i(t)$ can largely influence the probability of assigning $i$ to community $k$.
The above model (\ref{model:eq1}) also implies that, given $Z$, the entries of $A$ are conditional independent, and are further conditional independent of $X$.
Similar models were adopted in the literature of the SBM involving vector-valued nodal covariates; see \cite{weng2022}.
Hence, (\ref{model:eq1}) is an extension to the functional covariate setting.

\section{Variational Nonparametric Testing}
\label{sec:testing}

Denote $\theta=(\bB,\alpha,\beta)\in\Theta:=[0,1]^{(K+1)\times (K+1)}\times\bbR^K\times\cH_{m,K}$.
In practice, we are interested in testing the properties of $\theta$. For instance, we want to test if $\theta$ is equal to a hypothesized value, or test if $\beta$ can degenerate to a parametric form such as a polynomial.
These can be phrased as the following hypothesis testing problems:
\begin{equation}\label{simple:hypo}
    \textrm{(simple hypothesis)\,\,\,\,\,\,\,\,  $H_0: \theta=\theta^0$ vs. $H_1: \theta\neq\theta^0$,}
\end{equation}
where $\theta^0:=(\bB^0,\alpha^0,\beta^0)\in\Theta$ is a hypothesized value of $\theta$, or
\begin{equation}\label{composite:hypo}
 \textrm{(composite hypothesis)\,\,\,\,\,\,\,\, $H_0: \beta\in \cP_l$ vs. $H_1: \beta\in\cH_{m,K}\backslash\cP_l$,}
\end{equation}
where $\cP_l$ with $l<m$ is the class of $\beta=(\beta_1,\ldots,\beta_K)$ in which $\beta_1,\ldots,\beta_K$ are $l$-th order polynomials on $[0,1]$. Hypotheses (\ref{simple:hypo}) are useful to investigate the significance of the functional covariate $X$, e.g., large magnitudes of $\beta^0_k$ typically indicate stronger impact of $X_i(t)$'s on the class probabilities. Hypotheses (\ref{composite:hypo}) are useful for model diagonosis, e.g., 
whether the nonparametric model could degenerate to a parametric one.
We shall propose computationally efficient test statistics for (\ref{simple:hypo})
and (\ref{composite:hypo}), respectively.

Under (\ref{model:eq1}), we get the following likelihood function of the complete graph model (CGM) analogous to \cite{bickel2013aos}:
\[
P(A,Z,X;\theta)\propto P(A|Z)P(Z|X)
=\prod_{i<j}B_{Z_iZ_j}^{A_{ij}}(1-B_{Z_iZ_j})^{1-A_{ij}}\prod_{i=1}^nW_{iZ_i}(\alpha,\beta).
\]
The CGM assumes that the assignment vector $Z$ is observable, and hence, the likelihood function is built upon the triple $A,Z,X$. The corresponding penalized negative log-likelihood is
\[
\ell_{n,\lambda}(A,Z,X;\theta)=-\log{P(A,Z,X;\theta)}+\frac{\lambda}{2}J(\beta),
\]
where
$J(\beta)=\sum_{k=1}^{\sang{K}}\int_0^1|\beta_k^{(m)}(t)|^2dt$ is the roughness penalty useful to regularize the smoothness of the estimation of $\beta$, and $\lambda>0$ is the smoothing parameter.
Note that when nodal covariates are vectors, the roughness penalty is not needed. 
Based on the CGM, one can estimate $\theta$ through minimizing $\ell_{n,\lambda}(A,Z,X;\theta)$ as follows:
\begin{equation}\label{MLE:CGM}
\widehat{\theta}^{CGM}_{n,\lambda}=\argmin_{\theta\in\Theta} \ell_{n,\lambda}(A,Z,X;\theta).
\end{equation}

In practice, $Z$ is often unobservable, hence, (\ref{MLE:CGM}) may not be applicable. A common strategy is to estimate $\theta$ using a variational method to be described below. 
First of all, let $P(A,X;\theta)$ denote the marginal distribution of $A,X$ under $\theta$.
For any probability measure $Q$ over $[K]^n$, the Kullback–Leibler divergence from $Q(Z)$ to $P(Z|A,X;\theta)$ is nonnegative, so we have
\[
\log P(A,X;\theta)=\max_{Q}\mathbb{E}_{Z\sim Q}\left[\log P(A,Z,X;\theta)-\log Q(Z)\right],
\]
where the maximum is taken over probability measures $Q$ over $Z\in[K]^n$.
Let $\mathcal{Q}$ be a collection of product measures:
\[
\mathcal{Q}=\left\{Q: Q(Z)=\prod_{i=1}^n q_{iZ_i},
\sum_{k=0}^K q_{ik}=1, q_{ik}\ge0\right\}.
\]
Then the variational MLE $\widehat{\theta}^{VAR}_{n,\lambda}$ is defined as
\begin{eqnarray}
\nonumber
\widehat{\theta}^{VAR}_{n,\lambda}&=&\argmin_{\theta\in\Theta}\min_{Q\in\mathcal{Q}}L_{n,\lambda}(Q;A,X;\theta)\\
\label{eq-varMLE}
&\equiv&\argmin_{\theta\in\Theta}\min_{Q\in\mathcal{Q}}\left[-\mathbb{E}_{Z\sim Q}\left\{\log P(A,Z,X;\theta)-\log Q(Z)\right\}+\frac{\lambda}{2} J(\beta)\right].
\end{eqnarray}
In literature, \cite{bickel2013aos} and \cite{weng2022} proposed variational MLEs when model parameters are finite-dimensional. Therefore, (\ref{eq-varMLE}) can be viewed as their functional extensions involving a roughness penalty to control the smoothness of the slope $\beta$. 
Recursive convex optimization algorithms could be easily performed when model parameters are finite-dimensional.
Nonetheless, it is not easy to directly adapt their algorithms here due to the infinite dimensionality of  $\beta$.
In Section \ref{sec:algarithms}, we will circumvent this difficulty by an RKHS technique coupled with the representer theorem.
In Section \ref{sec:theory}, we will show that the theoretical analysis of $\widehat{\theta}^{VAR}_{n,\lambda}$ could be fulfilled by the aid of $\widehat{\theta}^{CGM}_{n,\lambda}$, which is similar in spirit to \cite{bickel2013aos}.
But more involved techniques such as the contraction mapping are needed to deal with the infinite-dimensional term $\beta$.

Let $\widehat{Z}$ be a consistent estimate of $Z$.
We propose the following penalized likelihood ratio test for testing (\ref{simple:hypo}):
\begin{equation} \label{eq-PLRT}
\textrm{PLRT}_{n,\lambda}=\ell_{n,\lambda}(A,\widehat{Z},X;\theta^0)-
\ell_{n,\lambda}(A,\widehat{Z},X;\widehat{\theta}^{VAR}_{n,\lambda}).
\end{equation}
Let $\widehat{\theta}^{VAR,0}_{n,\lambda}$ be the minimizer of $\ell_{n,\lambda}(\theta)$ over $\theta\in\Theta^0:=[0,1]^{(K+1)\times(K+1)}\times\mathbb{R}^K\times\cP_l$.
We propose the following composite likelihood ratio test for testing (\ref{composite:hypo}):
\[
\textrm{PLRT}^{comp}_{n,\lambda}=\ell_{n,\lambda}(A,\widehat{Z},X;\widehat{\theta}^{VAR,0}_{n,\lambda})-\ell_{n,\lambda}(A,\widehat{Z},X;\widehat{\theta}^{VAR}_{n,\lambda}).
\]
Both $\textrm{PLRT}_{n,\lambda}$ and $\textrm{PLRT}^{comp}_{n,\lambda}$ only involve variational MLEs
$\widehat{\theta}^{VAR,0}_{n,\lambda}$ and $\widehat{\theta}^{VAR}_{n,\lambda}$, hence, are computationally efficient.
Detailed computations of $\widehat{Z}$ will be provided in Section \ref{sec:algarithms}.
As far as we know, these are the first variational nonparametric test statistics whose null distributions shall be provided in Section \ref{sec:theory}.

\section{Estimation Algorithms}\label{sec:algarithms}
In this section, we provide algorithms for estimating the community labels $Z$ and
a recursive updating algorithm to compute the variational MLE $\widehat{\theta}^{VAR}_{n,\lambda}$.

Let $\widehat{\cX}$ be the $n\times(K+1)$ matrix consisting of the leading $K+1$ eigenvectors of $A$
satisfying $\widehat{\cX}^\top\widehat{\cX}=I_{K+1}$.
Let $\overline{\cX}$ be the row normalization of $\widehat{\cX}$. Specifically, let
\[
\overline{\cX}_i=\frac{\widehat{\cX}_i}{\|\widehat{\cX}_i\|_2},\,\,i=1,\ldots,n,
\]
where $\widehat{\cX}_i$ is the $i$th row of $\widehat{\cX}$ and $\|\cdot\|_2$ denotes the Euclidean norm of a vector.
Then let $\overline{\cX}$ be an $n\times(K+1)$ matrix with rows $\overline{\cX}_1,\ldots,\overline{\cX}_n$.
Let $\cM_{n\times(K+1)}$ be the class of $n\times(K+1)$ matrices that have exactly $K+1$ distinct rows.
We develop two algorithms to find $\widehat{Z}$.
The first one is to find $S^*\in\cM_{n\times(K+1)}$ such that
\begin{equation}\label{k:means:alg}
\|S^*-\overline{\cX}\|_F\le\gamma\inf_{S\in\cM_{n\times(K+1)}}\|S-\overline{\cX}\|_F,
\end{equation}
where $\|\cdot\|_F$ denotes the Frobenius norm, and $\gamma\ge1$ is a fixed constant.
Note that (\ref{k:means:alg}) degenerates to the classic $k$-means algorithm when $\gamma=1$.
In general, $\gamma>1$ produces computationally more efficient results; see Section 2.5.3 of \cite{DG17PhD}.
We estimate $Z_1,\ldots,Z_n$ as follows. Let $S_i^*$ be the $i$th row of $S^*$.
For $i,j\in[n]$, let 
\begin{equation}\label{est:Z:eqn}
\textrm{$\widehat{Z}_i=\widehat{Z}_j$ if $S_i^*=S_j^*$.}
\end{equation}
The second algorithm is to find distinct $Q_1^*,\ldots,Q_{K+1}^*\in\RR^{K+1}$ such that
\begin{equation}\label{k:means:alg:alt}
\max_{i\in[n]}\min_{1\le l\le K+1}\|Q_l^*-\overline{\cX}_i\|_2\le\gamma\inf_{\textrm{distinct $\widetilde{Q}_1,\ldots,\widetilde{Q}_{K+1}\in\RR^{K+1}$}}\max_{i\in[n]}\min_{1\le l\le K+1}\|\widetilde{Q}_l-\overline{\cX}_i\|_2,
\end{equation}
where $\gamma\ge1$ is constant.
For $i\in[n]$, let $1\le l_i\le K+1$ such that 
\[
\|Q_{l_i}^*-\overline{\cX}_i\|_2=\min_{1\le l\le K+1}\|Q_l^*-\overline{\cX}_i\|_2,
\]
and let $S^*\in\cM_{n\times(K+1)}$ whose $i$th row is $Q_{l_i}^*$.
Then we construct $\widehat{Z}$ similar to (\ref{est:Z:eqn}).
In practice, (\ref{k:means:alg}) may be computationally easier than (\ref{k:means:alg:alt}),
but the latter is proven strongly consistent which validates the proposed testing procedures;
see Proposition \ref{prop:weak:consistency:label}.

Next we compute $\widehat{\theta}^{VAR}_{n,\lambda}$. 
For notation convenience, let $\eta = (\alpha,\beta)$ with $\alpha=(\alpha_1, \dots, \alpha_K)\in\bbR^K$
and $\beta=(\beta_1, ,\ldots, \beta_K) \in\cH_{m,K}$.
For $r\ge0$, given the estimate of $Q(Z)$ at the $r$th iteration denoted by $q_{ik}^{(r)}$, we are solving
\begin{align} \label{eq-etaupdating}
\begin{split}
    \eta^{(r + 1)} & = \argmin_{\eta}   \sum_{i = 1}^n \left[-\sum_{k = 1}^K q_{ik}^{(r)}\left\{\alpha_k + \int_0^1 X_i(t)\beta_k(t) dt \right\}\right.\\
   & \qquad  \left. + \log \left\{1 + \sum_{k = 1}^K e^{\alpha_k + \int_0^1 X_i(t) \beta_k(t) dt}\right\}\right] + \lambda \sum_{k = 1}^K \int_0^1 |\beta_k^{(m)}(t)|^2dt. 
\end{split}
\end{align}
We resort to the representer theorem to solve the minimization problem of \eqref{eq-etaupdating}. Let $\psi_1, \ldots, \psi_m$ denote an orthonormal basis of the null space of $J$, i.e., $\{\beta: J(\beta) =  0\}$, and $K_1$ denote the reproducing kernel of its orthogonal complement. This reproducing kernel induces a linear operator from $L^2[0, 1]$ to $L^2[0, 1]$ such that $(K_1f)(t) = \int_0^1 K_1(s, t) f(s) ds$ for any $f \in L^2[0, 1]$.
With a slight modification of Theorem 1 of \cite{yuan2010}, we can easily show that each $\beta_k^{(r + 1)}$ admits the following representation:
\begin{equation} \label{eq-representer}
\beta_k^{(r + 1)}(t) = \sum_{j = 1}^m d_{jk}\psi_j(t) + \sum_{i = 1}^n c_{ik} (K_1X_i)(t)
\end{equation}
for some $\mbf d_k = (d_{1k}, \ldots, d_{mk})^\top \in \bbR^m$ and $\mbf c_{k} = (c_{1k}, \ldots, c_{nk})^\top \in \bbR^n$. Plugging \eqref{eq-representer} into \eqref{eq-etaupdating} yields 
\begin{equation} \label{eq-d&cupdating}
\begin{split}
    (\alpha^{(r+1)}, \sbf d^{(r + 1)}, \sbf c^{(r+1)}) = \argmin_{\mbf d \in \bbR^{Km}, \mbf c_k \in \bbR^{kn}}   \sum_{i = 1}^n \left\{-\sum_{k = 1}^K q_{ik}^{(r)}(\alpha_k + \mbf d_k^\top \mbf s_i + \mbf c_k^\top \mbf \xi_i)\right.\\
    \left. + \log \left(1 + \sum_{k = 1}^K e^{\alpha_k + \mbf d_k^\top \mbf s_i + \mbf c_k^\top \mbf \xi_i}\right)\right\} + \lambda \sum_{k = 1}^K \mbf c_k^\top \mbf \Xi c_k^\top,
\end{split}
\end{equation}
where $\mbf s_i = (s_{i1}, \ldots, s_{im})^\top$ with $s_{ij} = \int_0^1 X_i(t) \psi_j(t) dt$, $\mbf \xi_i = (\xi_{i1}, \ldots, \xi_{in})^{\top}$ with\\ $\xi_{ij} = \int_0^1\int_0^1 X_i(s) K_1(s, t) X_j(t) ds dt$ and $\mbf \Xi = (\mbf \xi_1, \ldots, \mbf \xi_n) \in \bbR^{n \times n}$. To facilitate the calculation, we consider applying the regularized reweighted least squares to solve the minimization problem of \eqref{eq-d&cupdating}. With slight abuse of notation, let $\eta_k = (\alpha_k, \sbf d_k^\top, \sbf c_k^\top) ^ \top$ and $\eta = (\eta_1^\top, \ldots, \eta_K^\top)^\top$. We set the score equations of \eqref{eq-d&cupdating}
\begin{equation} \label{eq-scoreIRLS}
S(\eta_k) = \sum_{i = 1}^n \{-q_{ik}^{(r)} + p_k(X_i(t); \eta) \} (1, \mbf s_i^\top, \xi_i^\top)^\top + 2\lambda (0, 0_m,  c_k^\top \mbf \Xi)^\top
\end{equation}
to be 0 for $k = 1, \ldots, K$. To solve the estimating equation \eqref{eq-scoreIRLS}, we employ the Newton-Raphson algorithm, which requires the Hessian matrix
$$
\frac{\partial S(\eta_k)}{\partial \eta_k} = -\sum_{i = 1}^n p_k(X_i(t); \eta) (1, \mbf s_i^\top, \xi_i^\top)^\top (1, \mbf s_i^\top, \xi_i^\top) + 2\lambda \cdot \diag(0, 0, \mbf \Xi). 
$$
Starting with $\eta^{(r)}$, a single Newton update is 
$$
\eta_k^{\mbox{new}} = \eta^{\mbox{old}} - \left(\frac{\partial S(\eta_k)}{\partial \eta_k}\right)^{-1} S(\eta_k^{\mbox{old}}),
$$
where the derivatives are evaluated at $\eta_k^{\mbox{old}}$ for $k = 1, \ldots, K$. Then $\eta^{(r+1)}$ is taken as the limit of this sequence when the Newton updates converge; it in turn provides the solution to \eqref{eq-etaupdating}. As illustrated in Chapter 4.4 of \cite{hastie2009}, the above Newton-Raphson update is equivalent to solving a weighted lease squares problem with an adjusted response. 

The updated probability parameter in the Bernoulli distribution is given by
%\bB \in \bbR^{(K + 1) \times (K + 1)}, 
\begin{equation*} 
    \bB^{(r + 1)} = \argmax_{\bB^\top = \bB} \sum_{a,b = 0}^K \left\{\log B_{ab} \cdot \sum_{i < j} A_{ij} q_{ia}^{(r)}q_{jb}^{(r)} + \log(1 - B_{ab}) \cdot \sum_{i < j} (1 - A_{ij}) q_{ia}^{(r)}q_{jb}^{(r)}\right\}.
\end{equation*}
This update has an analytic expression given by 
$$
B_{ab}^{(r + 1)} = \frac{\sum_{i < j} A_{ij} q_{ia}^{(r)}q_{jb}^{(r)}}{\sum_{i < j} q_{ia}^{(r)}q_{jb}^{(r)}}.
$$
More details about the derivation can be found in \cite{weng2022}. Then we update the distribution $Z$ as
\begin{align*}
\begin{split}
\{q_{ik}^{(r + 1)}\} = \argmax_{q_{ik}} \sum_{a,b=0}^{K} \left\{\log B_{ab}^{(r + 1)} \cdot \sum_{i < j} A_{ij} q_{ia} q_{jb}  + \log(1 - B_{ab}^{(r+1)}) \cdot \sum_{i < j} (1 - A_{ij}) q_{ia}q_{jb}\right\} \\
+ \sum_{i = 1}^n \sum_{k = 1}^K q_{ik} \left\{\alpha_k^{(r + 1)} + \int_0^1 X_i(t)\beta^{(r+1)}_k(t) dt\right\} - \sum_{i = 1}^n \sum_{k = 0}^K q_{ik} \log q_{ik}.
\end{split}
\end{align*}
As recommended in \cite{weng2022}, we employ an inner blockwise coordinate ascent method to solve it. Specifically, we update $\{q_{ik}\}_k$ one at a time:
\begin{align*}
\begin{split}
\{q_{ik}\}_k = \argmax_{\{q_{ik}\}_k} \sum_{k = 0}^K q_{ik} \left[\sum_b \sum_{j \neq i} \{A_{ij}q_{jb} \cdot \log B_{kb} + (1 - A_{ij})q_{jb} \cdot \log( 1- B_{kb})\}\right] \\
+ \sum_{k = 1}^K q_{ik} \left\{\alpha_k + \int_0^1 X_i(t)\beta_k(t) dt\right\} - \sum_{k = 0}^K q_{ik} \log q_{ik}. 
\end{split}
\end{align*}
Simple algebra leads to
$
q_{ik} = (\sum_{k = 0}^K e^{a_k})^{-1}e^{a_k}$, where
$$
a_k = \alpha_k + \int_0^1 X_i(t)\beta_k(t) dt + \sum_b \sum_{j \neq i}q_{jb} \{A_{ij}q_{jb} \cdot \log B_{kb} + (1 - A_{ij})q_{jb} \cdot \log( 1- B_{kb})\},
$$
with $\alpha_0 = \beta_0 = 0$. 

The following algorithm details the iteration steps, and it is called Variational functional Stochastic Block Model (to be abbreviated as VfSBM) in the article. The proposed estimation algorithm involves an important tuning parameter: smoothing parameter $\lambda$ in the penalty term, which controls the trade-off between fidelity to the data and smoothness of the estimated slope functions. We propose a cross-validation based method to choose $\lambda$. More specifically, we first randomly divide the entire nodes into $M$ folds and each time we hold out one fold as the test set. Then we implement Algorithm \ref{alg:var-fSBM} to the remaining $M - 1$ folds of nodes to estimate the slope functions. Then we estimate the membership for the test set and calculate the log-likelihood for the test set. Lastly, for a set of candidate values of $\lambda$, we apply the one standard error rule to the log-likelihood obtained from the test set to select the largest $\lambda$ among the candidate values. 
\begin{algorithm}
\caption{Solving \eqref{eq-varMLE} via iterating between $(\eta, \bB)$ and $Q$.} \label{alg:var-fSBM}
\begin{algorithmic}
\State Input: initialize $\{q^{(0)}_{ik}\}$, number of iterations $\mathcal{T}$ 
\State For $r =0,\dots, \mathcal{T}-1$
\begin{itemize}
\item[](a) update $\eta^{(r+1)}$ via solving \eqref{eq-d&cupdating} by the regularized reweighted least squares
\item[](b) update $B^{(r+1)}_{ab}=\frac{\sum_{i< j} A_{ij} q^{(r)}_{ia}q^{(r)}_{jb}}{\sum_{i< j} q^{(r)}_{ia}q^{(r)}_{jb}}$
\item[](c) Update $\{q_{ik}^{(r+1)}\}$ via Repeating 
\begin{itemize}
\item[] For $i=1,\dots, n$
\begin{itemize}
\item[] $\log q_{ik} \propto \alpha_k^{(r+1)} + \int_0^1 X_i(t)\beta_k^{(r+1)}(t) dt + \sum_b \sum_{j \neq i}q_{jb} \{A_{ij}q_{jb} \cdot \log B_{kb}^{(r + 1)} + (1 - A_{ij})q_{jb} \cdot \log( 1- B_{kb}^{(r + 1)})\}$ 
\end{itemize}
\end{itemize}
\end{itemize}
\State Output: $\{q_{ik}^{(\mathcal{T})}\}, \eta^{(\mathcal{T})}, \bB^{(\mathcal{T})}$.
\end{algorithmic}
\end{algorithm}

\section{Asymptotic Analysis}\label{sec:theory}
In this section, we shall derive the rate of convergence for $\widehat{\theta}^{VAR}_{n,\lambda}$,
derive consistency for $\widehat{Z}$, and derive the asymptotic null distribution for the penalized likelihood ratio tests.

Let $\theta^0=(\bB^0,\alpha^0,\beta^0)$ denote the unknown true value of $\theta$
under which $A,Z,X$ are generated through (\ref{model:eq1}).
For $i\in[n]$ and $k\in[K]$, define $R_i(\alpha_k,\beta_k)=\alpha_k+\int_0^1X_i(t)\beta_k(t)dt$.
Throughout we shall write $\eta=(\alpha,\beta)$ for simplification.
It is easy to see that the penalized likelihood based on CGM is
\begin{eqnarray*}
\ell_{n,\lambda}(A,Z,X;\theta)
&=&-\sum_{i<j}\left[A_{ij}\log{B_{Z_iZ_j}}+(1-A_{ij})\log{(1-B_{Z_iZ_j})}\right]\\
&&-\sum_{i=1}^n\left[I(Z_i\neq 0)R_i(\alpha_{Z_i},\beta_{Z_i})-\log\left(1+\sum_{k=1}^{K}e^{R_i(\alpha_k,\beta_k)}\right)\right]+\frac{\lambda}{2}J(\beta)\\
&\equiv&\ell(A,Z,X;\theta)+\frac{\lambda}{2}J(\beta).
\end{eqnarray*}
The Fr\'{e}chet derivatives of $\ell_{n,\lambda}(\theta):=\ell_{n,\lambda}(A,X,Z;\theta)$ w.r.t. $\theta$ are summarized as follows: for $\Delta\theta,\Delta\theta^{(1)},\Delta\theta^{(2)}\in\Theta$,
\begin{eqnarray*}
S_{n,\lambda}(\theta)\Delta\theta&\equiv&D\ell_{n,\lambda}(\theta)\Delta\theta\\
&=&-\sum_{i<j}\left(\frac{A_{ij}}{B_{Z_iZ_j}}-\frac{1-A_{ij}}{1-B_{Z_iZ_j}}\right)\Delta B_{Z_iZ_j}-\sum_{i=1}^n(I_i-W_i(\eta))^\top R_i(\Delta\alpha,\Delta\beta) \\
&&+\lambda J(\beta,\Delta\beta),
\end{eqnarray*}
\begin{eqnarray*}
DS_{n,\lambda}(\theta)\Delta\theta^{(1)}\Delta\theta^{(2)}
&=&\sum_{i<j}\left(\frac{A_{ij}}{(B_{Z_iZ_j})^2}+\frac{1-A_{ij}}{(1-B_{Z_iZ_j})^2}\right)\Delta B_{Z_iZ_j}^{(1)}\Delta B_{Z_iZ_j}^{(2)}\\
&&+\sum_{i=1}^nR_i(\Delta\beta^{(1)})^\top\boldsymbol{\Sigma}_{i}(\beta)R_i(\Delta\beta^{(2)})
+\lambda J(\Delta\beta^{(1)},\Delta\beta^{(2)}),
\end{eqnarray*}
where
$I_i=(\II(Z_i=1),\ldots,\II(Z_i=K))^\top$,
$R_i(\eta)=(R_i(\alpha_1,\beta_1),\ldots,R_i(\alpha_K,\beta_K))^\top$, and $W_i(\eta)=(W_{i1}(\eta),\ldots,W_{iK}(\eta))^\top$ with
$W_{ia}(\eta)$ defined in (\ref{eqn:Wia}),
$\boldsymbol{\Sigma}_{i}(\eta)=\textrm{diag}(W_{i}(\eta))-W_{i}(\eta)W_{i}(\eta)^\top\in\bbR^{K\times K}$.
For $a\in[K]$, let $\pi_a=P(Z=a)=\mathbb{E}_X\left[W_a(\theta^0)\right]$.
It is easy to verify that
\begin{eqnarray*}
&&S_\lambda(\theta)\Delta\theta\\
&\equiv&\bE S_{n,\lambda}(\theta)\Delta\theta\\
&=&\binom{n}{2}\sum_{a,b=0}^K\frac{\pi_a\pi_b(B_{ab}-B_{ab}^0)}{B_{ab}(1-B_{ab})}\Delta B_{ab}+n\sum_{a=1}^K\bE_X\left[(W_a(\eta)-W_a(\eta^0))R(\Delta\eta_a)\right]+\lambda J(\beta,\Delta\beta)
\end{eqnarray*}
and
\begin{eqnarray*}
DS_\lambda(\theta)\Delta\theta^{(1)}\Delta\theta^{(2)}
&=&\binom{n}{2}\sum_{a,b=0}^K\pi_a\pi_b\left[\frac{B_{ab}^0}{(B_{ab})^2}+\frac{1-B_{ab}^0}{(1-B_{ab})^2}\right]\Delta B^{(1)}_{ab}\Delta B^{(2)}_{ab}\\
&&+n\E_X R(\Delta\eta^{(1)})^\top\boldsymbol{\Sigma}(\eta) R(\Delta\eta^{(2)})+\lambda J(\Delta\beta^{(1)}),\Delta\beta^{(2)})).
\end{eqnarray*}
We make some technical assumptions.
\begin{assumption}\label{A2}
There exists some $c>0$ such that, almost surely,
\[
c^{-1}\le\lambda_{\min}(\bs{\Sigma}(\eta^0))\le\lambda_{\max}(\bs{\Sigma}(\eta^0))\le c.
\]
Moreover, for any $t\in[0,1]$,
$\bE_X\left[X(t)\boldsymbol{\Sigma}(\eta^0)\right]=0$.
\end{assumption}

Let $\boldsymbol{\Omega}=\bE_X\boldsymbol{\Sigma}(\eta^0)$
and $\bC(s,t)=\bE_X\left[\boldsymbol{\Sigma}(\eta^0)X(s)X(t)\right]$, $s,t\in[0,1]$. Here, $\bC(s,t)$
maps any $(s,t)\in[0,1]\times[0,1]$ to a $ K \times K$ matrix.
For any $\theta=(\bB,\eta)$, $\widetilde{\theta}=(\widetilde{\bB},\widetilde{\eta})\in\Theta$,
define
\begin{eqnarray*}
\langle\theta,\widetilde{\theta}\rangle&=&
DS_\lambda(\theta^0)\theta\widetilde{\theta}\\
&=&\binom{n}{2}\sum_{a,b=0}^K\frac{\pi_a\pi_b}{B^0_{ab}(1-B^0_{ab})}B_{ab}\widetilde{B}_{ab}+n\alpha^\top\boldsymbol{\Omega}\widetilde{\alpha}+n\int_0^1\int_0^1
\beta(s)^\top\bC(s,t)\widetilde{\beta}(t)dsdt\\
&&+\lambda J(\beta,\widetilde{\beta}).
\end{eqnarray*}
Define
\[
V(\beta,\widetilde{\beta})=\int_0^1\int_0^1\beta(s)^\top \bC(s,t) \widetilde{\beta}(t)dsdt,
\]
and for any $\beta,\widetilde{\beta}\in\cH_{m,K}$, define  
\begin{equation}\label{inner:prod}
\langle\beta,\widetilde{\beta}\rangle_1=V(\beta,\widetilde{\beta})+\frac{\lambda}{n} J(\beta,\widetilde{\beta}).
\end{equation}

\begin{assumption}\label{A1}
For any $\beta\in\mathcal{H}_{m,K}$, $V(\beta,\beta)=0$ if and only if $\beta=0$.
\end{assumption}

\begin{assumption}\label{A2.5} There exist non-negative constants $\zeta,\kappa$ with $\zeta>\kappa+1/2$ such that $\rho_\nu\asymp \nu^{2\zeta}$ and $\|\varphi_\nu\|_\infty\le C_\varphi\nu^{\kappa}$ for all $\nu\ge1$.
\end{assumption}

\begin{assumption}\label{A3} 
There exist constants $\tau, C>0$ such that $\bE\exp(\tau\|X\|_{L^2})<\infty$ and, for any $\beta\in H^m[0,1]$,
			\[
			\E\left\{\left|\int_0^1 X_i(t)\beta(t)dt\right|^4\right\} \le C \left[   \E\left\{\left|\int_0^1 X_i(t)\beta(t)dt\right|^2\right\}              \right]^2.
			\label{eq-moment}
			\]
\end{assumption}

\begin{assumption}\label{A4} 
There exists $\rho_n\in(0,1)$ with $\rho_n=o(1)$ and $\log{n}=o(n\rho_n)$ such that $\bB^0=\rho_n S$, where $S\in[0,1]^{(K+1)\times (K+1)}$ is symmetric and has no identical columns.
\end{assumption}

\begin{remark}\label{rem}
Assumption \ref{A2} requires the Fisher information matrix $\bs{\Sigma}(\eta^0)$ being invertible, and $X(t)$ being properly centered to ensure identifiability. Similar conditions are useful to derive a valid inner product of the RKHS in nonparametric literature; see \cite{shang2015aos}. 
Assumption \ref{A1} requires the bilinear form $V$ being positive definite, which is equivalent to the positive definiteness of the covariance kernel $\bC$.
Assumption \ref{A2.5} requires growth rates for $\rho_\nu$ and the supremum norm of $\varphi_\nu$.
Specifically, the supremum norm of the $\nu$-th eigenfunction $\|\varphi_\nu\|_\infty$ diverges at a polynomial order $\nu^\kappa$ for a constant $\kappa\ge0$, and the $\nu$-th eigenvalue $\rho_\nu$ diverges at a polynomial order $\nu^{2\zeta}$.
A special case is that the covariance kernel $\bC$ satisfies pseudo Sacks–Ylvisaker conditions of order $\varrho\ge0$, and then Assumption \ref{A2.5} holds with $\kappa=\varrho+1$ and $\zeta=m+\varrho+1$; see Proposition 2.2 of \cite{shang2015aos}. 
Assumption \ref{A3} requires an exponential tail condition for $\|X\|_{L^2}$ and a fourth-moment condition for $X$, which is common in functional data literature; see \cite{yuan2010,cai2012jasa} and \cite{shang2015aos}.
Assumption \ref{A4} is standard in network data analysis which requires the true symmetric probability matrix $bB^0$
being decreasing at rate $\rho_n=o(1)$.
We require the columns of $S$ being nonidentical and $\rho_n\gg\log{n}/n$, which are necessary for consistently estimating the vertex labels even in conventional stochastic block models without nodal information; see \cite{bickel2009pnas,zhao2012consistency}. 
\end{remark}
\begin{lemmain}\label{lem:RKHS}
Under Assumption \ref{A1}, $(\mathcal{H}_{m,K},\langle\cdot,\cdot\rangle_1)$ is an RKHS.
\end{lemmain}
\begin{lemmain}\label{simultaneous:diag}
Under Assumptions \ref{A2} and \ref{A1}, 
there exist $\varphi_{\nu}\in\cH_{m,K}$ and $\rho_{\nu}>0$, for $\nu\ge1$, such that
$V(\varphi_{\nu},\varphi_{\mu})=\delta_{\nu\mu}$ and $J(\varphi_{\nu},\varphi_{\mu})=\rho_{\nu}\delta_{\nu\mu}$,
where $\delta_{\nu\mu}$ is Kronecker's notation.
\end{lemmain}
Lemma \ref{lem:RKHS} guarantees that $\cH_{m,K}$ is an RKHS under the inner product $\langle\cdot,\cdot\rangle_1$.
Lemma \ref{simultaneous:diag} guarantees that bilinear forms $V$ and $J$ can be diagonalized by the same eigen-pairs $(\rho_\nu,\varphi_\nu),\nu\ge1$.
Let $d(\widehat{Z},Z)=\sum_{i=1}^n \II(\widehat{Z}_i\neq Z_i)$ be the clustering error, also viewed as a distance between $\widehat{Z}$ and $Z$. 
\begin{proposition}\label{prop:weak:consistency:label}
Suppose Assumptions \ref{A3} and \ref{A4} are fulfilled. 
\begin{itemize}
\item[(1).] (Weak consistency) With probability approaching one,
$d(\widehat{Z},Z)\le C_1/\rho_n$, where $C_1$ is an absolute constant depending on $K,\gamma$,
and $\widehat{Z}$ is based on (\ref{k:means:alg}).
\item[(2).] (Strong consistency) Moreover, if $n\rho_n\gg(\log{n})^\psi$ for a constant $\psi>2$, then $\PP(\widehat{Z}=Z)\to1$ as $n\to\infty$,
where $\widehat{Z}$ is based on (\ref{k:means:alg:alt}).
\end{itemize}
\end{proposition}
Under a weaker assumption $n\rho_n\gg\log{n}$, strong consistency of spectral clustering has been also derived by \cite{Su2020IT} in the conventional stochastic block model by using a recursive contraction mapping 
technique to control the perturbation between empirical and population eigeivectors. 
To reduce technical involvement, we adopt a more direct method to prove strong consistency,
which uses a tight perturbation bound on eigenvectors (in $\ell_\infty$-norm) by \cite{pmlr-v83-eldridge18a}.
Though our proof is shorter, it requires a stronger condition $n\rho_n\gg(\log{n})^\psi$ for $\psi>2$.

Theorem \ref{main:thm:2} establishes the convergence rate for $\theta^{CGM}_{n,\lambda}$ as well as a Bahadur representation (\ref{eqn:cgm:bahadur}).
Let $h=(\lambda/n)^{\frac{1}{2\zeta}}$, where $\zeta$ is given in Assumption \ref{A2.5}.
\begin{theorem}\label{main:thm:2}
Under Assumptions \ref{A2}-\ref{A3}, and $2\zeta>6\kappa+3$,
$h=o(1)$, $nh^{6\kappa+4}\to\infty$.
Then $\|\widehat{\theta}_{n,\lambda}^{CGM}-\theta^0\|=O_\Pr(r_n)$, where $r_n=\lambda^{1/2}+h^{-1/2}$.
Moreover, we have
\begin{equation}\label{eqn:cgm:bahadur}
\|\widehat{\theta}_{n,\lambda}^{CGM}-\theta^0+S_{n,\lambda}(\theta^0)\|
=O_\Pr\left(\frac{r_n^2}{nh^{4\kappa+2}}+\frac{r_n^4}{nh^{6\kappa+3}}+\frac{r_n^2}{\sqrt{n}}\right).
\end{equation}
\end{theorem}
Let $\textrm{per}(\theta^0)$ be the number of permutations $\sigma:[K]\mapsto[K]$ such that $\sigma(\theta^0)=\theta^0$.
Theorem \ref{main:thm:4} establishes a Bahadur representation (\ref{eqn:var:bahadur}) for $\widehat{\theta}^{VAR}_{n,\lambda}$ when $per(\theta^0)=1$. Comparing (\ref{eqn:cgm:bahadur}) with (\ref{eqn:var:bahadur}),
we conclude that $\widehat{\theta}^{VAR}_{n,\lambda}$ has the same high-order approximation, e.g., $S_{n,\lambda}(\theta^0)$, as $\widehat{\theta}^{CGM}_{n,\lambda}$.
\begin{theorem}\label{main:thm:4}
Under the assumptions of Theorem \ref{main:thm:2}, if $per(\theta^0)=1$ and
$\|\widehat{\theta}_{n,\lambda}^{VAR}\|=O_\Pr(\sqrt{nh^{2\kappa+1}})$, then we have
\begin{equation}\label{eqn:var:bahadur}
\|\widehat{\theta}_{n,\lambda}^{VAR}-\theta^0+S_{n,\lambda}(\theta^0)\|
=O_\Pr\left(\frac{r_n^2}{nh^{4\kappa+2}}+\frac{r_n^4}{nh^{6\kappa+3}}+\frac{r_n^2}{\sqrt{n}}\right).
\end{equation}
\end{theorem}
Equations (\ref{eqn:cgm:bahadur}) and (\ref{eqn:var:bahadur}) provide Bahadur representations for the 
CGM and variational MLEs, respectively.
The RHS of each equation involves an order depending on $r_n$ and $h$, which is usually faster than $r_n$.
Specifically, the optimal convergence rate for $\widehat{\theta}_{n,\lambda}^{CGM}$ and $\widehat{\theta}_{n,\lambda}^{VAR}$ is $r_n=n^{\frac{1}{2(2\zeta+1)}}$ which is achieved at $\lambda=n^{\frac{1}{2\zeta+1}}$. 
When $\kappa=1$ and $\zeta=m+1$ (See Remark \ref{rem} about the validity of this scenario), 
the order of RHS becomes $\max\{n^{\frac{-2m+8}{2\zeta+1}},n^{\frac{0.5-\zeta}{2\zeta+1}}\}$
which is faster than $r_n$ if $m>15/4$. If we further require $m>9/2$, then the RHS becomes $o(n^{-\frac{1}{2\zeta+1}})$ which is sufficient to remove asymptotic bias from $\widehat{\beta}^{VAR}_{n,\lambda}$.
As a consequence, we have the following pointwise asymptotic normality of $\widehat{\beta}^{VAR}_{n,\lambda}$.
\begin{corollary}\label{main:cor}
For any $t\in[0,1]$, as $n\to\infty$, $\sqrt{nh}\left(\widehat{\beta}^{VAR}_{n,\lambda}(t)-\beta^0(t)\right)\overset{d}{\to}N(0,\sigma_t^2)$,
where 
\[
\textrm{$\sigma_t^2=\lim_{h\to0}\sum_{\nu=1}^\infty\frac{h}{(1+h^{2\zeta}\rho_\nu)^2}\varphi_\nu(t)\varphi_\nu(t)^\top$ finitely exists.}
\]
\end{corollary}

The following theorem provides null asymptotic distribution for the variational test statistic $\textrm{PLRT}_{n,\lambda}$ defined in \eqref{eq-PLRT}. 
\begin{theorem}\label{main:thm:5}
Under the assumptions of Theorem \ref{main:thm:4}, if $2\zeta>6\kappa+3$ and $h\asymp n^{-c}$ with
\begin{equation}\label{main:thm:5:rate:conditions}
\max\left\{\frac{4}{16\zeta-24\kappa-11},\frac{2}{4\zeta+1}\right\}<c<\frac{1}{6\kappa+5},
\end{equation}
and $n\rho_n\gg(\log{n})^\psi$ for a constant $\psi>2$, then under $H_0$ in (\ref{simple:hypo}), as $n\to\infty$,
\[
\sqrt{\frac{2h}{m_2}}\left(\textrm{PLRT}_{n,\lambda}-\frac{m_1}{2h}\right)\overset{d}{\to}N(0,1),
\]
where $m_p=\sum_{\nu=1}^\infty\frac{h}{(1+h^{2\zeta}\rho_\nu)^p}$ for $p\ge1$.
\end{theorem}
Theorem \ref{main:thm:5} derives the asymptotic distribution of $\textrm{PLRT}_{n,\lambda}$ under $H_0$
when $h\asymp n^{-c}$ for $c$ satisfying (\ref{main:thm:5:rate:conditions}).
This theorem yields an approximate null distribution 
\[
\frac{2m_1}{m_2}\times\textrm{PLRT}_{n,\lambda}\overset{a}{\sim}\chi^2_{\frac{m_1^2}{m_2h}}.
\]
A simple consequence of Theorem \ref{main:thm:5} is the consistency of penalized likelihood ratio test for composite hypothesis (\ref{composite:hypo}).
More precisely, under $H_0$ in (\ref{composite:hypo}), the true parameter $\theta^0$
has parametric component $\beta^0$, hence, conventional likelihood theory guarantees that $\ell_{n,\lambda}(\widehat{\theta}^{VAR,0})-\ell_{n,\lambda}(\theta^0)=O_\Pr(1)$. This implies that
\[
\sqrt{\frac{2h}{m_2}}\left[\textrm{PLTR}^{comp}_{n,\lambda}-\frac{m_1}{2h}\right]=\sqrt{\frac{2h}{m_2}}\left[\ell_{n,\lambda}(\theta^0)-\ell_{n,\lambda}(\widehat{\theta}^{VAR})-\frac{m_1}{2h}\right]+O_\Pr(\sqrt{h})\overset{d}{\to}N(0,1).
\]
Therefore, $\textrm{PLTR}^{comp}_{n,\lambda}$ admits the same asymptotic null distribution as the PLRT test for a simple hypothesis.
Implementation of the penalized likelihood ratio test entails an estimation of $\rho_{\nu}$'s, which can be estimated  
by {\bf Chebfun} function in Matlab following the idea of \cite{hao2021semiparametric}.
Specifically, by Proposition 2.2 of \cite{shang2015aos}, it is sufficient to find an estimated $\rho_{\nu}$ through solving the following differential equations:
\begin{align} \label{eq-eigenfunction}
\begin{cases}
(-1)^m y_{\nu}^{(2m)}(t) = \rho_{\nu} \int_0^1 C(s, t)y_{\nu}(s) ds, \\
y_{\nu}^{(j)}(0) = y_{\nu}^{(j)}(1)= 0, ~~~j = m, \ldots, 2m - 1.
\end{cases}
\end{align}
If $(\rho_{\nu}, y_{\nu})$ denotes the eigenvalue and the eigenfunction of \eqref{eq-eigenfunction}, then $\varphi_{\nu} = y_{\nu}/\sqrt{V(y_{\nu}, y_{\nu})}$ and $\rho_{\nu}$ satisfy Lemma \ref{simultaneous:diag}.

\begin{comment}
\begin{figure}[H]
\centering
\includegraphics[width=0.7\textwidth]{Xprofile.pdf}
\caption{The trajectories of $X$ from 10 randomly selected nodes.}
\label{fig:Xprof}
\end{figure}
\end{comment}

\section{Simulation Studies} \label{sec:simulation}
In this section, we demonstrate the finite-sample performance of the proposed  VfSBM algorithm, as outlined in Algorithm \ref{alg:var-fSBM}, through simulation studies.

\subsection{Data Generation}
We generate $n$ nodes, and the nodewise functional covariates are independently generated by $X_i(t) = \sum_{j = 1}^5 \zeta_j \xi_{ij} \phi_j(t)$, where $\zeta_j = (-1)^{j + 1} j^{-2}$, $\xi_{ij}$'s are i.i.d uniformly distributed on $[-\sqrt{3}, \sqrt{3}]$, and the eigenfunctions satisfy $\phi_1(t) = 1$ and for $j \geq 1$, $\phi_j(t) = \sqrt{2}\cos{(j-1)\pi t}$ for $t \in [0, 1]$. 
The functional covariates are observed at 50 time points on [0, 1] for all nodes. 
% Figure \ref{fig:Xprof} displays the trajectories of $X$ from 10 randomly selected nodes. 
We take the number of communities to be 2. 
Then the community membership of each node, $Z_i$, is independently generated by the logistic regression: 
\begin{equation} \label{eq-datagen}
P(Z_i = 1 \mid  X_i) = \frac{\exp\left\{\alpha^0 + \int_0^1 \beta^0(t) X_i(t) dt\right\}}{ 1+ \exp\left\{\alpha^0 + \int_0^1 \beta^0(t) X_i(t) dt\right\}},
\end{equation}
where $\alpha^0 = 0.1$ and $\beta^0(t) = 50(t - 0.5)^2 - 5$. To generate the $n \times n$ adjacency matrix $\bA = (A_{ij})$, we consider the design in \cite{weng2022}. More specifically, 
given the community memberships of nodes $i$ and $j$ ($i < j$), the (symmetric) binary edge between them, $A_{ij}$, is generated from the Bernoulli distribution with $\bB = \rho_n \begin{bmatrix}
    1.2 & 0.3 \\
   0.3  & 1.2
\end{bmatrix}$ with $\rho_n = (\log n)^{1.5}/n$. To better assess the performance of the proposed method, 500 independent simulations were run. Additionally, we present simulation results for various values of $\lambda$ and $n$: $\lambda \in \{10^{-2}, 10^{-3}, 10^{-4}, 10^{-5}\}$ and $n \in \{100, 120, 140, \ldots, 400\}$.

We compare Algorithm \ref{alg:var-fSBM} with two alternative methods: the vanilla SBM and the spectral clustering method  in terms of community detection.
%Note we need to vary the number of nodes $n$ and $\tau$ in $\tau \beta^0$.} 
In the vanilla SBM (to be abbreviated VSBM), we only take the generated adjacency matrix $\bA$ into consideration when estimating the community structure. 
In terms of spectral clustering, we treat each functional covariate as a 50-dimensional covariate and then apply the 
assortative covariate-assisted spectral clustering (to be abbreviated as SPEC) proposed by \cite{binkiewicz2017}. To implement the proposed VfSBM algorithm, we take the clustering result from the SPEC as the initial value of $q_{ik}$ in ALgorithm \ref{alg:var-fSBM}, and employ the tuning parameter selection method proposed in Section \ref{sec:algarithms} to select $\lambda$. We employ the commonly used metric to evaluate the community detection performances of these two methods: Normalized Mutual Information (NMI in short) (\citealp{ana2003}), which is defined as
$$
\mbox{NMI} = \frac{-2 \sum_i \sum_j n_{ij} \log \left(\frac{n_{ij} n}{n_{i\cdot}n_{\cdot j}}\right)}{\sum_i n_{i\cdot} \log \left(\frac{n_{i\cdot}}{n}\right) + \sum_j n_{\cdot j} \log \left(\frac{n_{\cdot j}}{n}\right)}.
$$
In the above expression, $n_{i\cdot}$ denotes the number of nodes in community $i$, $n_{\cdot j}$ denotes the number of nodes whose estimated label is $j$ and $n_{ij}$ denotes the number of nodes whose true label is $i$ but estimated one is $j$. The value of NMI is bounded by 1 from above and a larger value indicates better community recovery. 

\subsection{Performances}
\begin{figure}[ht]
\centering
\includegraphics[width=0.9\textwidth]{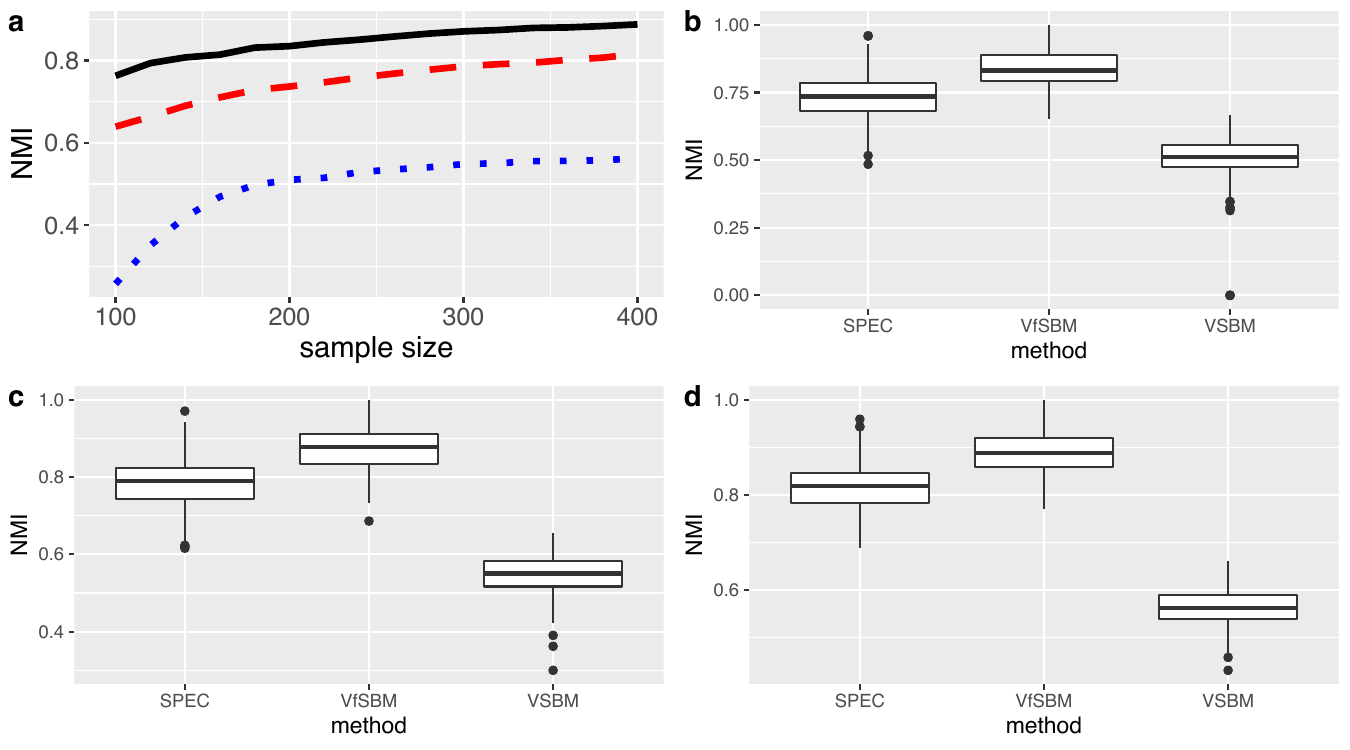}
\caption{(a) The average NMIS of the three methods across 500 simulation runs for various sample sizes, where we take $\lambda = 10^{-4}$ when implementing our algorithm. The solid black line, the red dashed line and the blue dotted line represent the average NMIs for VfSBM, SPEC and VSBM, respectively. (b), (c) \& (d): the boxplots of NMIs of VfSBM for $n = 200, 300$ and 400, respectively.}
\label{fig:NMI4}
\end{figure}

Since most often our algorithm chooses $\lambda = 10^{-4}$ regardless of the number of nodes, we present the numerical results for $\lambda = 10^{-4}$. Panel (a) in Figure \ref{fig:NMI4} displays the average NMIs obtained from the three methods across the 500 simulation runs. In contrast to the curse of dimensionality that the spectral clustering method may suffer, the proposed functional data based method benefits from the dense observations of the trajectory. 
By leveraging the smoothness properties of the trajectory of the functional covariate, the proposed method shows a remarkable advantage over the spectral clustering method. To compare the NMI obtained from different methods, we show the boxplots of the NMIs obtained from them for $n = 200, 300$ and 400 in panels (b), (c) and (d). Obviously our method outperforms the two competitors regardless of the number of nodes. Moreover, as the number of nodes diverges, the NMI of our method approaches to 1, indicating a near prefect recovery of the network structure.

\begin{figure}[H]
\centering
\includegraphics[width=0.7\textwidth]{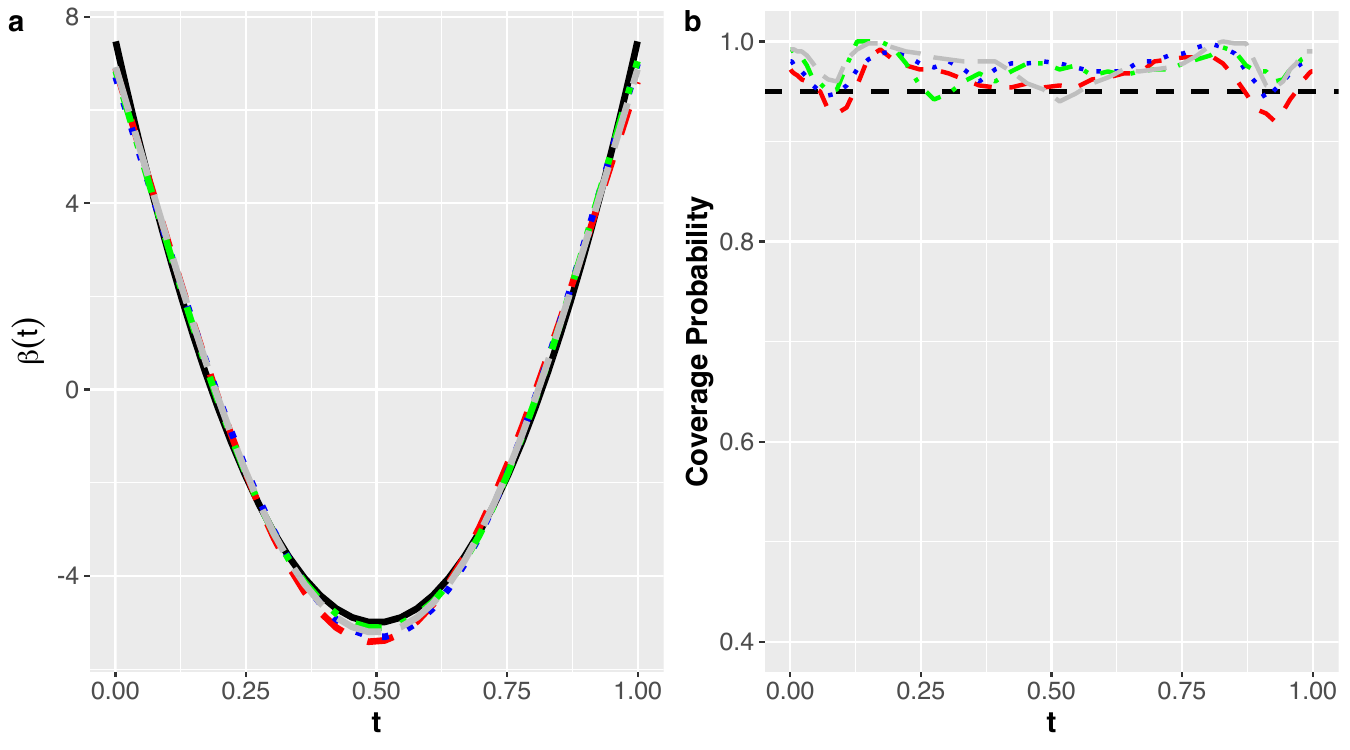}
\caption{(a) The average of the estimated $\beta$ across the 500 simulation runs for various sample sizes: $n \in \{100, 200, 300, 400\}$, where we take \sang{$\lambda = 10^{-4}$} when implementing our algorithm. The solid black line depicts the true $\beta_0$. (b) The true coverage probabilities of the pointwise confidence intervals constructed from the estimated $\beta$ from the 500 simulation runs with the same sample sizes as panel (a). The horizontal black dashed represents the nominal level: 0.95. In both panels, the read dashed, the blue dotted, the green dot-dash and the grey long-dash lines represent the estimates from $n = 100, 200, 300$ and 400, respectively. }
\label{fig:beta4}
\end{figure}

%and the method by \cite{weng2022}
Besides the community detection performance, we also investigate the performance in parameter estimation for the proposed VfSBM. Panel (a) in Figure \ref{fig:beta4} depicts the pointwise average of the estimated $\beta(t)$ across the 500 simulation runs with $n \in \{100, 200, 300, 400\}$ and $\lambda = 10^{-4}$. It demonstrates that the VfSBM yields a reasonable estimate of the slope function that reflects the impact of the nodal functional covariate on community detection. Additionally, as the number of nodes grows, the estimated slope function becomes closer to the true function; it justifies the consistency property of the proposed variational estimator established in Theorem \ref{main:thm:4}. 

\begin{table}[H]
\centering

\addtolength{\tabcolsep}{3pt}    
\begin{tabular}{ccccc}
	\hline
	\multirow{2}{*}{$t$}&
	\multicolumn{4}{c}{$n$}    \\
	\cline{2-5} 
	& 100 & 200  & 300 & 400  \\
	\cline{1-5}
	0.07	& 0.932 & 0.946 & 0.950 & 0.962    \\
	%\cline{2-11}
	%	\hline
	0.48	& 0.954 & 0.980 & 0.978  & 0.952    \\
	%\cline{2-11}
	%	\hline
	0.93	& 0.920 & 0.952 & 0.960 & 0.954    \\
	%\cline{2-11}
	%	\hline
	
	\hline
	
\end{tabular}

%\vspace*{5mm}

\caption{
	Summary of coverage probabilities of the pointwise confidence intervals for $\beta^0(t)$ at $t = 0.07, 0.48$ and 0.93 with $n \in \{100, 200, 300, 400\}$ nodes.}	

\label{tab:CP}
\end{table}

We further construct pointwise confidence intervals based on Corollary \ref{main:cor}. 
%Note that $S_{n, \lambda}(\theta^0)$ can be written as the sum of some i.i.d random functions with mean 0. Then we can establish a (pointwise) weak convergence for $\hat{\beta}(t)$ by resorting to the Lindeberg central limit theorem and taking a suitable normalizing constant. 
The main problem is to estimate the standard error of $\hat{\beta}(t)$ for each $t \in [0, 1]$. To tackle this issue, we resort to the regularized weighted least square formulation in Algorithm \ref{alg:var-fSBM}. By replacing the exponential part in the logistic regression by its quadratic approximation to formulate the weighted least squares problem, one gets a Gaussian likelihood with the adjusted responses and variances. Then the results of Chapter 3.3 in \cite{gu2013} give an approximate posterior variances for $\hat{\beta}(t)$, which is then used to construct pointwise confidence intervals for $\beta^0(t)$. 
Panel (b) in Figure \ref{fig:beta4} shows that the true coverage probability of pointwise confidence intervals constructed from the variational estimate across 500 simulation runs as in panel (a).
Table \ref{tab:CP} further details the coverage probabilities of the pointwise confidence intervals for $\beta^0(t)$ at $t = 0.07, 0.48$ and 0.93 with various node sizes. Figure \ref{fig:beta4} and Table \ref{tab:CP} demonstrate that the true coverage probabilities of our proposed pointwise confidence intervals are close to the nominal level when the number of nodes is reasonably large.

\

We also conduct simulations for $\lambda \in \{10^{-2}, 10^{-3}, 10^{-5}\}$, and the results are summarized in Figures \ref{fig:NMI2} to \ref{fig:beta5} in the supplementary material. These results showcase the effect of $\lambda$ when implementing our variational algorithm. When the selected $\lambda$ is too large, excessive shrinkage would be induced. Consequently, the estimated slope function has a relatively large bias, and this would has a negative effect on the coverage probabilities of the pointwise confidence intervals, as indicated in Figures \ref{fig:NMI2} and \ref{fig:beta2} . On the other hand, Figures \ref{fig:NMI3} to \ref{fig:beta5} demonstrate that our variational  method performs well in both recovering the network structure and estimating the effect of nodal information over a wide range of $\lambda$. 

We further study the performance of the penalized likelihood ratio test in finite samples. In particular, we choose $\alpha^0 = 0$ and $\beta^0(t) = \{50(t - 0.5)^2 - 5\} \cdot  \tau$ in model \eqref{eq-datagen} when generating $n = \{100, 200, 300, 400\}$ nodes, and $\tau = \{0, 0.4, 0.6, 0.8, 1\}$ to examine the size and power of the test for composite hypothesis \eqref{composite:hypo}. In each scenario, the rejection rate is calculated at 5\% significance level based on 500 independent simulation runs.
Table \ref{tab:PLRTrej} summarizes the size and the power of the penalized likelihood ratio test in each simulation scenario. We find that the size of the test is always smaller than the significance level, while its power increases as $\tau$ increases when the null hypothesis does not hold and the number of nodes is fixed. Furthermore, when $\tau \neq 0$ is fixed, the power of our test increases with respect to the number of nodes; this demonstrates the consistency of the proposed test.

\begin{table}[H]
	\centering
	
	\addtolength{\tabcolsep}{3pt}    
	\begin{tabular}{cccccc}
		\hline
		\multirow{2}{*}{$n$}&
		\multicolumn{5}{c}{$\tau$}    \\
		\cline{2-6} 
		& 0  & 0.4 & 0.6 & 0.8 & 1 \\
		\cline{1-6}
		100	& 0.040 & 0.573 & 0.637 & 0.701 & 0.814   \\
		%\cline{2-11}
		%	\hline
		200	& 0.024  & 0.612 & 0.778 & 0.876 &0.928  \\
		%\cline{2-11}
		%	\hline
		300	& 0.026& 0.656 & 0.820 & 0.936 &0.966 \\
	400	& 0.034 & 0.714 & 0.908 & 0.954 & 0.970 \\
	%	500	& 0.012 & 0.180 & 0.422 & 0.610 & 0.772 & 0.842 \\
		\hline
		
	\end{tabular}
	
	%\vspace*{5mm}
	
	\caption{
		Summary of rejection probabilities of the PLRT at 5\% significance level for different values of $\tau$ with $n \in \{100, 200, 300, 400\}$ nodes.}	
	
	\label{tab:PLRTrej}
\end{table}

\section{Real Applications} 
\label{sec:application}

\subsection{MyFitnessPal Data}
We illustrate our method using the MyFitnessPal dataset \citep{weber2016insights}, which contains 587,187 days of food diary records inputted by 9.9K MyFitnessPal users from September 2014 through April 2015. 
The original dataset was retrieved from 
\url{https://www.kaggle.com/datasets/zvikinozadze/myfitnesspal-dataset}.
Then it was pre-processed according to the pipeline at \url{https://github.com/RRaphaell/MyFitnessPal} for our study purpose. 
% to get the matrix \verb|myFitnessPal_parsed.csv| from the raw file \verb|mfp-diaries.tsv|. 
% The matrix 
The pre-processed dataset records the daily food intake logged by the user, including the food names, actual intake calories, and planned calories.

Out of all the food dairy records, we take complete observations of 67 days starting from 2015-01-05 (Monday) and ending with 2015-03-12 (Thursday) and restricting to 921 out of total 9,896 participants. In our network analysis, each node represents one of these 921 subjects. The dataset contains $(0,1)$ valued similarity scores between any pair of persons, which constitute a  similarity matrix $S = (s_{ij}) \in \mathbb{R}^{921 \times 921}$. Similarity is defined by the commonality of the lifestyles such as food habits, and it is defined as the mean Token Set Ratio between the list of strings containing the names of food that two users have taken on the same date. The Token Set Ratio measures the string similarity, and can be implemented in the Python library \texttt{FuzzyWuzzy}. Then the binary adjacency matrix is defined as follows: $A_{ij} = 1$ if the $s_{ij} \geq m$, where $m$ denotes the median of all $s_{ij}$'s, and 0 otherwise. 
Additionally, the dataset collected the daily intake calories and  planned calories  of each person during the study period, which are denoted as $X_1(t)$ and $X_2(t)$, respectively.

\begin{figure}[ht]
\centering
\includegraphics[width=0.7\textwidth]{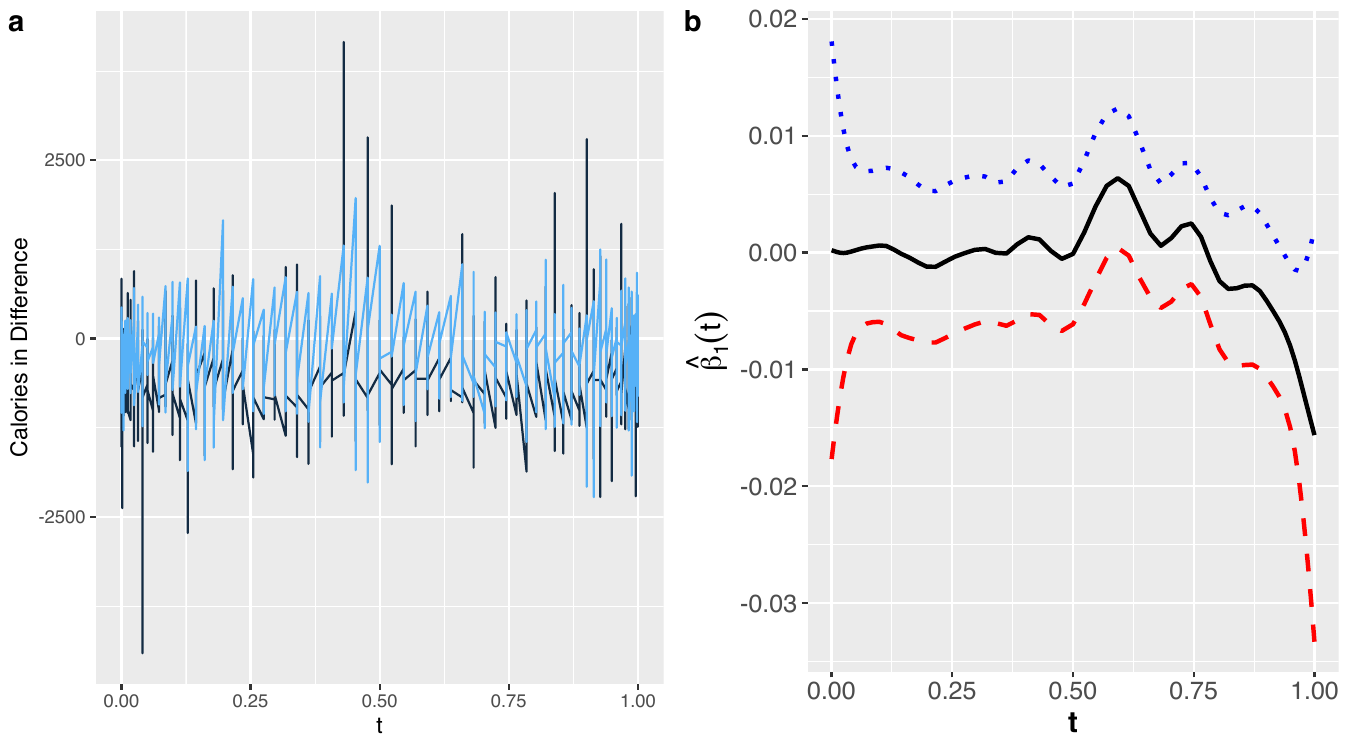}
\caption{(a) The trajectory of the difference in intake and planned calories from 2015-01-05 to 2015-03-12 for 20 randomly selected participants of the two groups. (b) The estimated slope functions along with 95\% pointwise confidence intervals of the trajectory of the difference in intake and planned calories in models \eqref{eqn:Wia} and \eqref{model:eq1} for the food diary data.}
\label{fig:calorie-beta}
\end{figure}
\cite{weber2016insights} categorized all participants into two groups, ``above" and ``below", based on whether their daily consumption is higher than the ``daily calories goal". Furthermore, they found that logging food with ``oil", ``butter" or ``mcdonalds" in the name indicates exceeding one's calories goals. As we define the adjacency matrix through the similarity score based on the list of strings containing the names of food, we choose the number of communities to be 2 in the subsequent analysis, i.e., $K = 1$.
In light of the classification rule in \cite{weber2016insights}, we take the difference in the daily intake and planned calories, $X(t) = X_1(t) - X_2(t)$, as the functional nodal information as it can reflect whether the real consumption exceeds the daily calories goals for each participant. Therefore, it must be predictive of the community label. We scale the time span to [0, 1] for convenience. 
After applying our proposed method to this dataset, we display the estimated slope function, $\hat{\beta}_1(t)$, as well as 95\% pointwise confidence intervals on the right panel of Figure \ref{fig:calorie-beta}. From the figure we find that the effect of $X$ is not significant in the first half of the study period, while a marked effect of $X$ is present in the second half. This finding coincides with the comparison between the intake calories and the planned ones, as displayed in the left panel of Figure \ref{fig:calorie-beta}. Only nuanced differences in these two calories can be found for $t < 0.5$ from the 20 randomly sampled participants of the two groups. In contrast, these two groups of curves are better separated for $t > 0.5$, which indicates a more remarkable difference in the food diary pattern between these groups. We further apply the penalized likelihood ratio test in \eqref{eq-PLRT} to test the null effect of $X$. The test statistic gives rise to a $p$-value less than $10^{-6}$, which indicates a significant effect of the trajectory of the differences in intake and planned calories in the network structure.

%\newpage
\subsection{International Trade Data}
The international trade data consist of the yearly amount of international trades between 58 countries from 1981 to 2000 \citep{westveld2011mixed}. In particular, both the amount of imports and exports were recorded between any two of these 58 countries. The international trade network in 1995 has been widely studied in the literature of network analysis; see \cite{saldana2017many}, \cite{hu2020corrected}, \cite{wang2023fast} and references therein. Besides the trade data, we also collected the yearly GDP since 1980 for each of these 58 countries. 

We apply our proposed method to this year's trade network where each node represents one country. 
Following \cite{saldana2017many} and \cite{wang2023fast}, we define the undirected binary network as follows. Let $W_{ij} = \text{Trade}_{ij} + \text{Trade}_{ji}$, where $\text{Trade}_{ij}$ denotes the export amounts from country $i$ to country $j$. Then we set $A_{ij} = 1$ if $W_{ij}$ is equal to or exceeds the median of all $W_{ij}$'s, and 0 otherwise. We choose the yearly GDP data from 1980 to 1995 to be the functional nodal information in models \eqref{eqn:Wia} and \eqref{model:eq1}. Actually, motivated by the gravity model, \cite{westveld2011mixed} proposed a longitudinal network where the logarithm of trades between country $i$ and country $j$ is regressed against the logarithm of the GDPs for both these countries, the geographical distance, and several additional covariates. Moreover, the community detection result by \cite{wang2023fast} mainly reflects the dissimilarities in both distance and the GPD level between difference clusters. These facts motivate us to account for the effect of the GDP trajectory in detecting the community structure in the international trade network.

\begin{table}[!t]
\textcolor{black}{\begin{tabular}{c|c}
\hline  Group & Countries \\\hline
1  &  \begin{tabular}[c]{@{}l@{}}Argentina, Australia, Austria, Belgium, Brazil, Canada, Chile, Denmark, \\
Finland, France, Germany, India, Ireland, Italy, Japan, South Korea, \\
Malaysia, Mexico, Netherlands, Norway, Portugal, Singapore, Spain, Sweden, \\
 Switzerland, Thailand, Turkey, United Kingdom, United States \end{tabular} \\\hline
2  &  \begin{tabular}[c]{@{}l@{}}Algeria, Colombia, Costa Rica, Ecuador, Egypt, EI Salvador, Greece, Guatemala \\
Indonesia, Israel, Morocco, New Zealand, Peru, Philippines, Trinidad and Tobago, Venezuela   \end{tabular}\\\hline
3  &  \begin{tabular}[c]{@{}l@{}}Barbados, Bolivia, Cyprus, Honduras, Iceland, Jamaica, Mauritius, Nepal, \\
Oman, Panama, Paraguay, Tunisia, Uruguay   \end{tabular} \\\hline
\end{tabular}
\caption{\label{tab:trade} Community detection result for the international trade data based on our method.}}
\end{table}

\begin{figure}[H]
\centering
\includegraphics[width=0.5\textwidth]{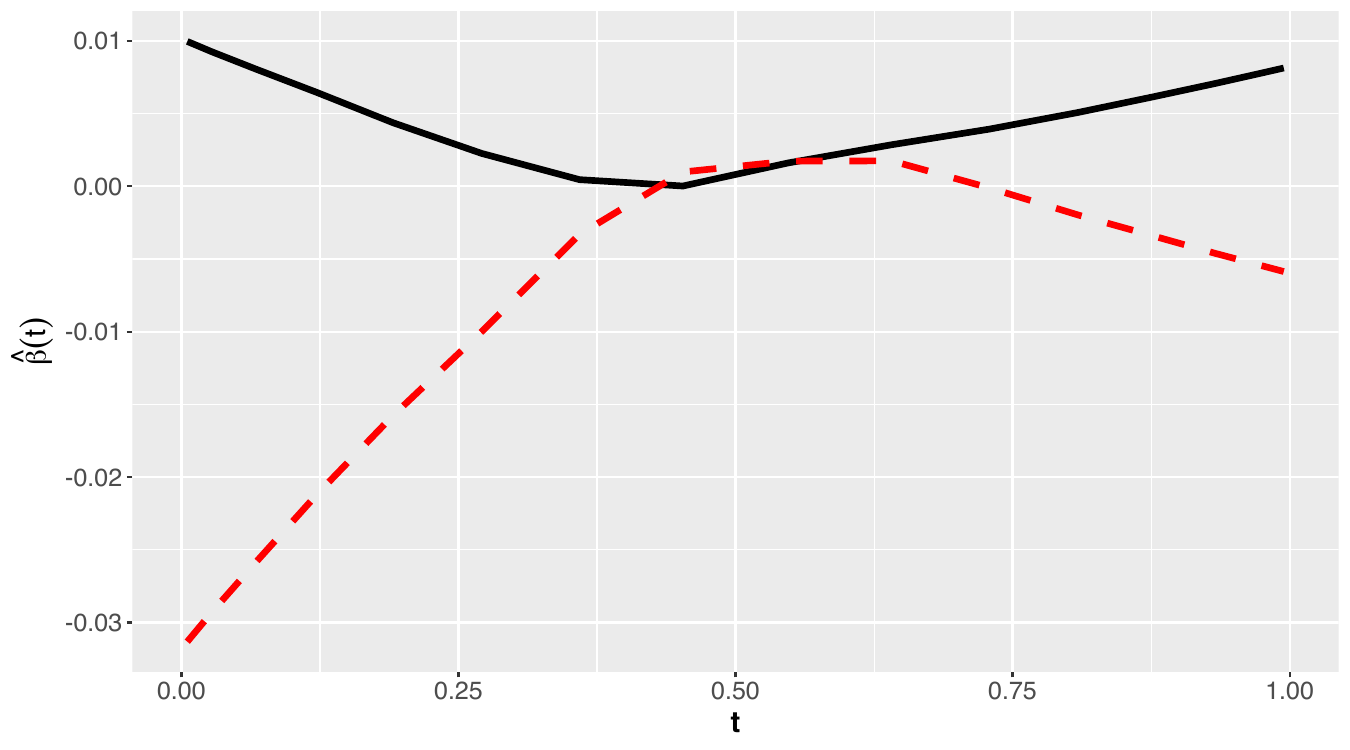}
\caption{The estimated slope functions of the GDP trajectory in models \eqref{eqn:Wia} and \eqref{model:eq1} for the international trade data. The solid red line and dashed red line represent $\hat{\beta}_1(t)$ and $\hat{\beta}_2(t)$, respectively.}
\label{fig:beta_trade}
\end{figure}

\cite{saldana2017many} and \cite{hu2020corrected} selected the number of the communities to be 3 after implementing their respective selection procedure. This conclusion was further confirmed by \cite{wang2023fast} with their PPL method at the significance level of 0.01. Therefore, we consider $K = 2$ in our analysis. 
Table \ref{tab:trade} displays the community detection result for the international trade network. Our analysis suggests that countries like Argentina, Australia and Austria should be put in the same group with developed countries. In contrast, \cite{wang2023fast} found that these countries should not be put in the same group. This difference reflects the role of the GDP trajectory in determining the community structure. Moreover, Figure \ref{fig:beta_trade} displays the estimated slope functions, $\hat{\beta}_1$ and $\hat{\beta}_2(t)$, for the GDP trajectory. Here we scale the time span to be [0, 1]. As argued in \cite{westveld2011mixed}, the geographical distance plays an important role in determining the network structure in early 1980's, but becomes less and less important. In contrast, the GDP level is always an important factor, which may be elusive in Figure \ref{fig:beta_trade}. Nevertheless, when we conduct a hypothesis testing to test whether the GDP trajectory has no effect on the network structure, i.e., $\beta_1 = \beta_2 = 0$, our proposed test statistic leads to a the 
$p$-value that is much smaller than 0.01. This result indicates a significant effect of the GDP trajectory in determining the network structure.

\section{Conclusions} \label{sec:conclusion}
In this paper, we propose a new model called fSBM that can jointly perform community detection and hypothesis testing in the presence functional covariates associated with network vertices.
Our method is variational, and hence, is computationally efficient. Besides, it has applications in real-world data sets such as GDP data and MyFitnessPal data, where the network vertices both involve functional covariates, and our method is able to perform valid community detection and statistical inferences.

The proposed fSBM is nonetheless in the scope of unsupervised learning since it doesn't involve response variables. In some applications such as MyFitnessPal data, an important aim is to predict future outcomes such as future fitness results, hence, a supervised learning extension of the current work might be useful. For instance, we plan to involve response variables in fSBM, and propose variational algorithms to jointly conduct community detection, hypothesis testing and prediction.   
\bibliography{arxiv} 
\bibliographystyle{natbib}

\newpage
\begin{center}
	Supplementary Material to the Manuscript Entitled ``Variational Nonparametric Testing in Functional Stochastic Block Model"
\end{center}
\setcounter{equation}{0}
\setcounter{figure}{0}
\setcounter{table}{0}
\setcounter{page}{1}
\makeatletter
\renewcommand{\theequation}{A\arabic{equation}}
\renewcommand{\thefigure}{S\arabic{figure}}
\renewcommand{\bibnumfmt}[1]{[S#1]}
\renewcommand{\citenumfont}[1]{S#1}

\setcounter{section}{0}
\renewcommand{\thesection}{S.\arabic{section}}

The supplementary material contains some additional numerical results. 
\section{Additional simulation results}
\begin{figure}[H]
	\centering
	\includegraphics[width=0.99\textwidth]{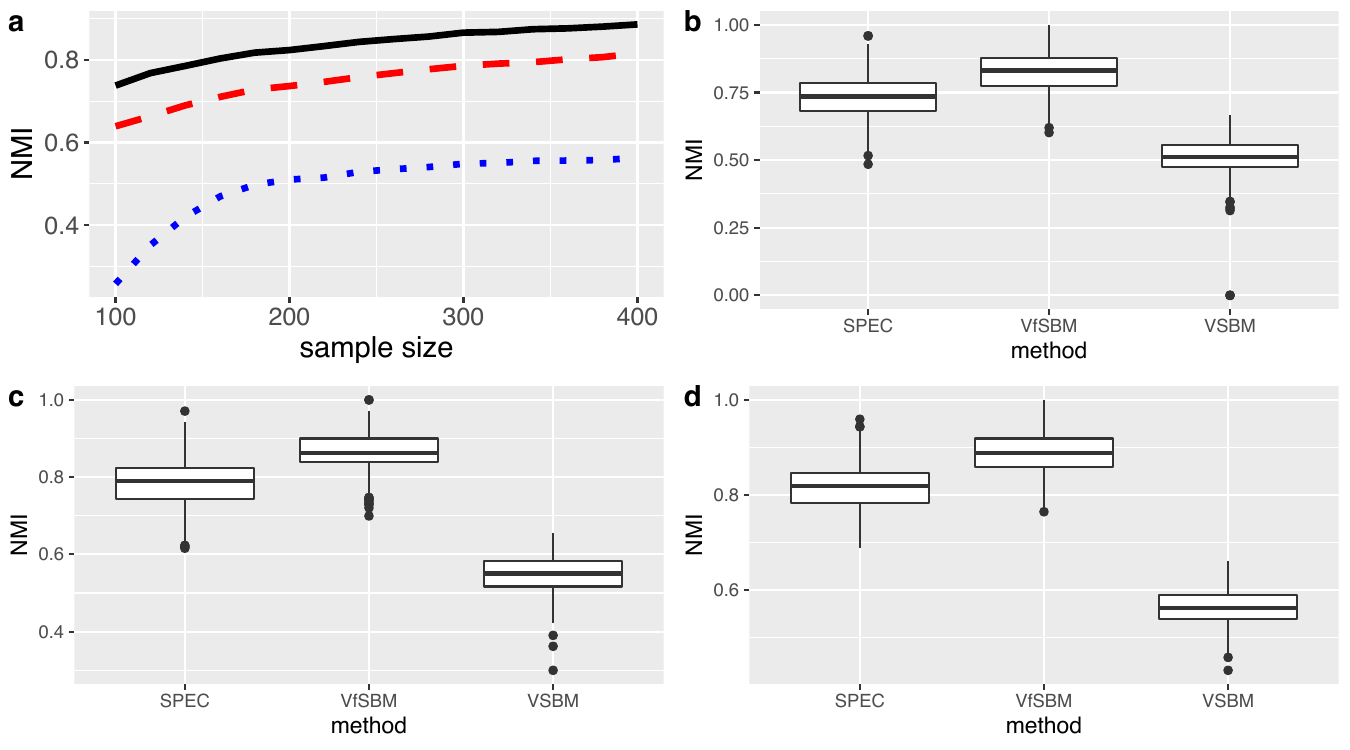}
	\caption{(a) The average NMIS of the three methods across 500 simulation runs for various sample sizes, where we take $\lambda = 10^{-2}$ when implementing our algorithm. The solid black line, the red dashed line and the blue dotted line represent the average NMIs for VfSBM, SPEC and VSBM, respectively. (b), (c) \& (d): the boxplots of NMIs of VfSBM for $n = 200, 300$ and 400, respectively. }
	\label{fig:NMI2}
\end{figure}

\begin{figure}[H]
	\centering
	\includegraphics[width=0.99\textwidth]{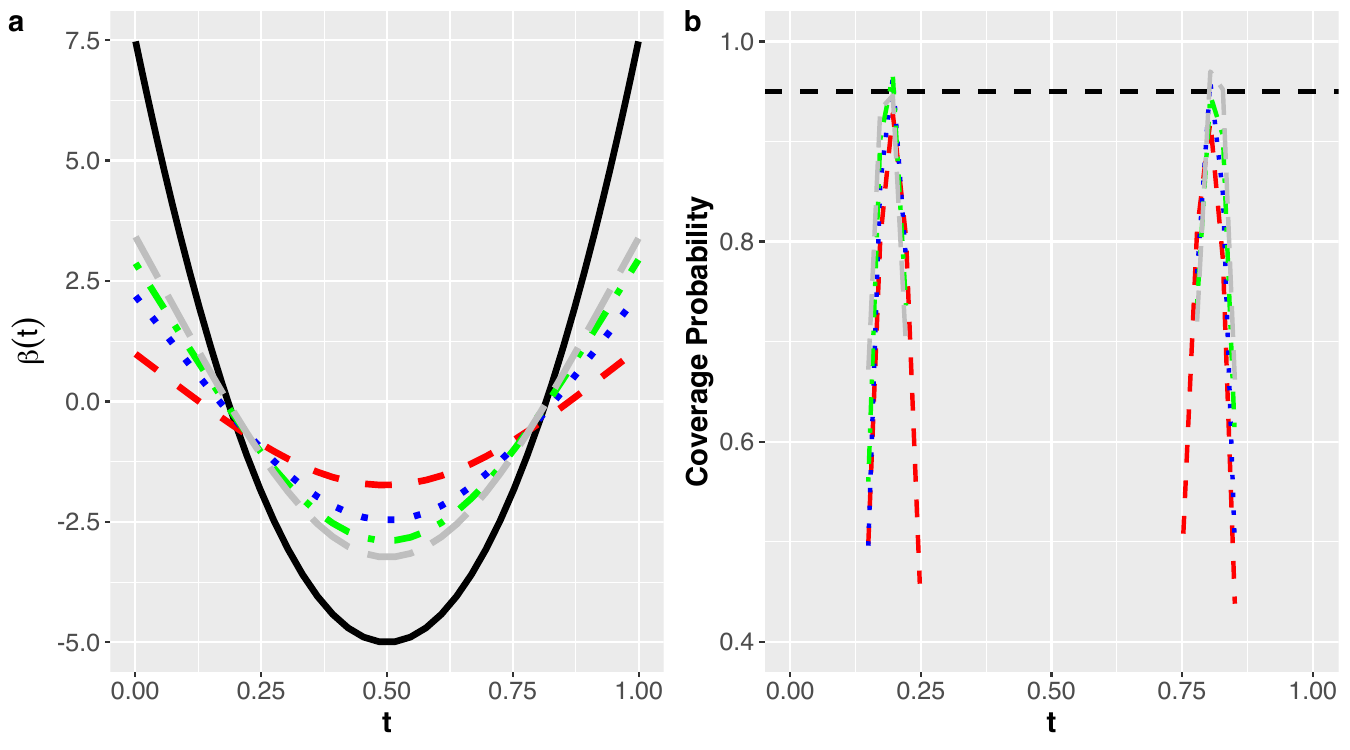}
	\caption{(a) The average of the estimated $\beta$ across the 500 simulation runs for various sample sizes: $n \in \{100, 200, 300, 400\}$, where we take $\lambda = 10^{-2}$ when implementing our algorithm. The solid black line depicts the true $\beta_0$. (b) The true coverage probabilities of the pointwise confidence intervals constructed from the estimated $\beta$ from the 500 simulation runs with the same sample sizes as panel (a). The horizontal black dashed represents the nominal level: 0.95. In both panels, the read dashed, the blue 
		%and the violet two-dash
		dotted, the green dot-dash and the grey long-dash  lines represent the estimates from $n = 100, 200, 300$ and 400, respectively. }
	\label{fig:beta2}
\end{figure}

\begin{figure}[H]
	\centering
	\includegraphics[width=0.99\textwidth]{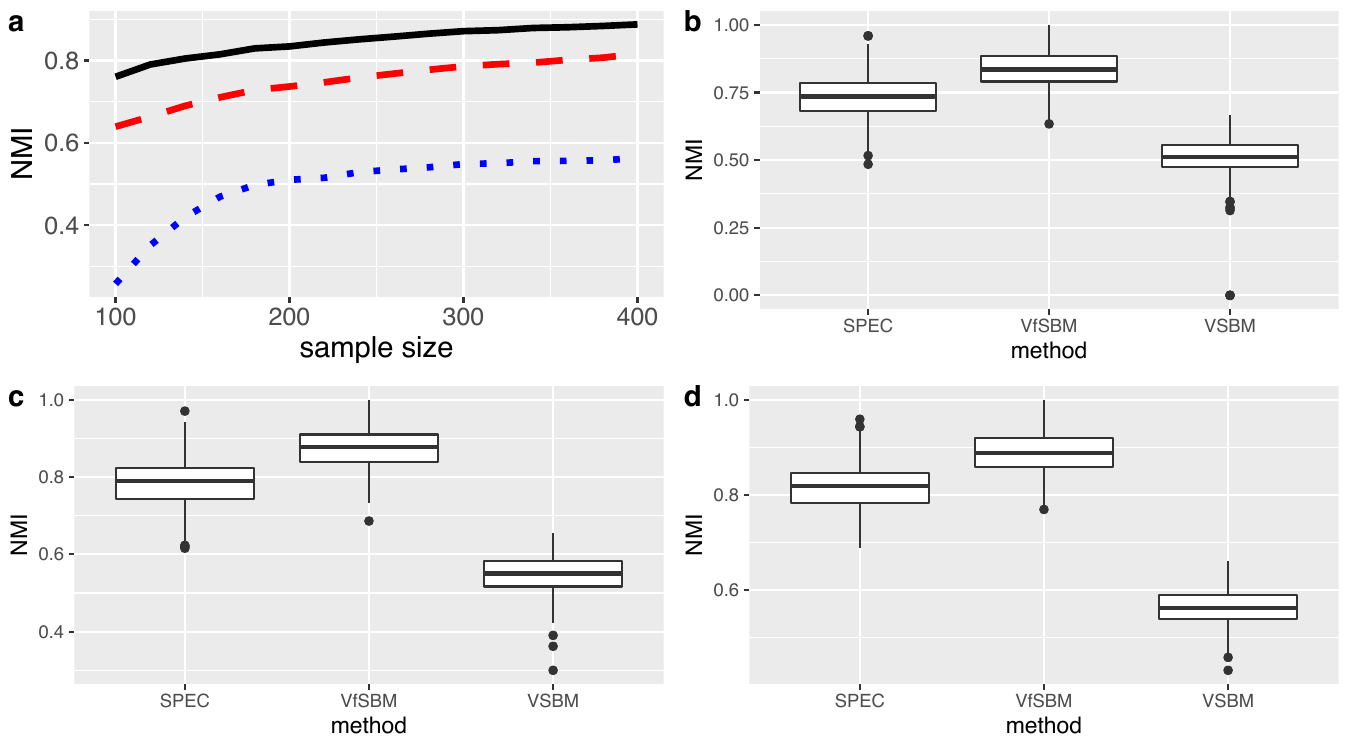}
	\caption{(a) The average NMIS of the three methods across 500 simulation runs for various sample sizes, where we take \sang{$\lambda = 10^{-3}$} when implementing our algorithm. The solid black line, the red dashed line and the blue dotted line represent the average NMIs for VfSBM, SPEC and VSBM, respectively. (b), (c) \& (d): the boxplots of NMIs of VfSBM for $n = 200, 300$ and 400, respectively. }
	\label{fig:NMI3}
\end{figure}

\begin{figure}[H]
	\centering
	\includegraphics[width=0.99\textwidth]{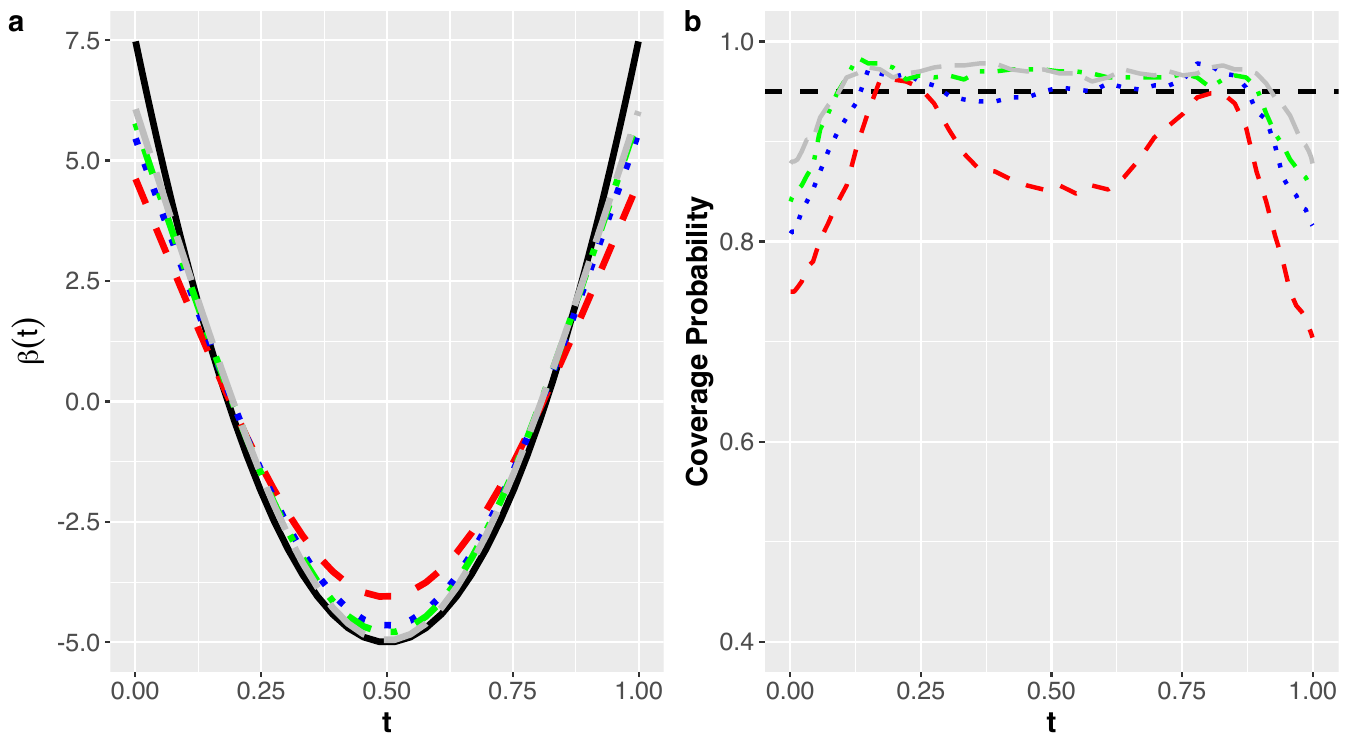}
	\caption{(a) The average of the estimated $\beta$ across the 500 simulation runs for various sample sizes: $n \in \{100, 200, 300, 400\}$, where we take \sang{$\lambda = 10^{-3}$} when implementing our algorithm. The solid black line depicts the true $\beta_0$. (b) The true coverage probabilities of the pointwise confidence intervals constructed from the estimated $\beta$ from the 500 simulation runs with the same sample sizes as panel (a). The horizontal black dashed represents the nominal level: 0.95. In both panels, the read dashed, the blue dotted, the green dot-dash and the grey long-dash lines represent the estimates from $n = 100, 200, 300$ and 400, respectively. }
	\label{fig:beta3}
\end{figure}

\begin{figure}[H]
	\centering
	\includegraphics[width=0.99\textwidth]{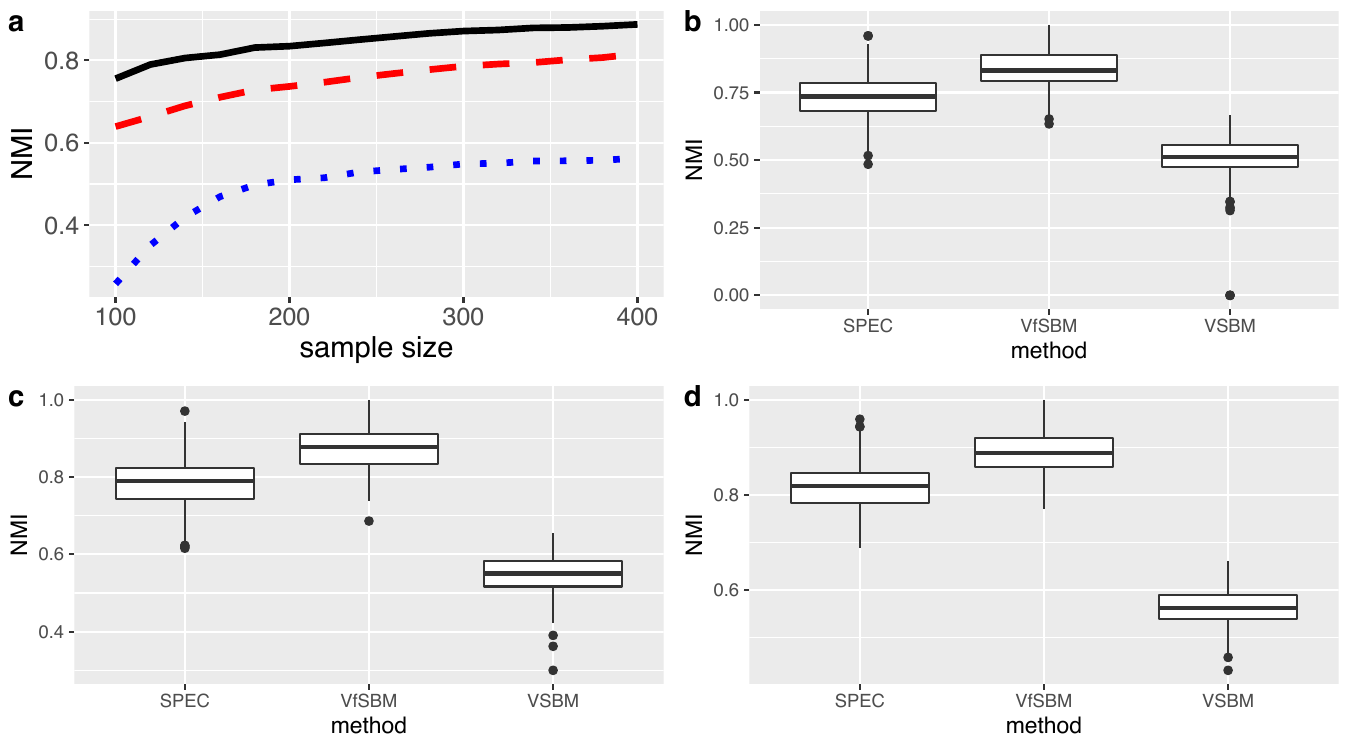}
	\caption{(a) The average NMIS of the three methods across 500 simulation runs for various sample sizes, where we take $\lambda = 10^{-5}$ when implementing our algorithm. The solid black line, the red dashed line and the blue dotted line represent the average NMIs for VfSBM, SPEC and VSBM, respectively. (b), (c) \& (d): the boxplots of NMIs of VfSBM for $n = 200, 300$ and 400, respectively. }
	\label{fig:NMI5}
\end{figure}

\begin{figure}[H]
	\centering
	\includegraphics[width=0.99\textwidth]{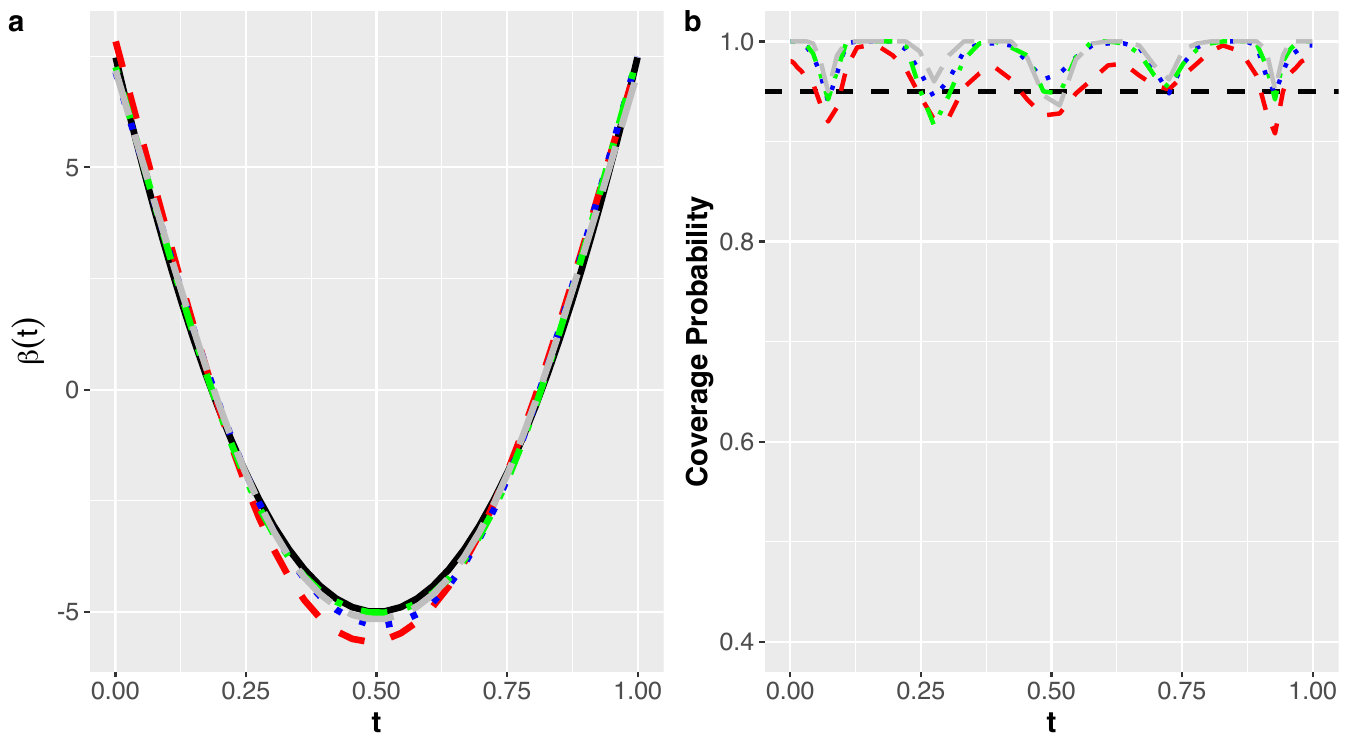}
	\caption{(a) The average of the estimated $\beta$ across the 500 simulation runs for various sample sizes: $n \in \{100, 200, 300, 400\}$, where we take $\lambda = 10^{-5}$ when implementing our algorithm. The solid black line depicts the true $\beta_0$. (b) The true coverage probabilities of the pointwise confidence intervals constructed from the estimated $\beta$ from the 500 simulation runs with the same sample sizes as panel (a). The horizontal black dashed represents the nominal level: 0.95. In both panels, the read dashed, the blue dotted, the green dot-dash and the grey long-dash lines represent the estimates from $n = 100, 200, 300$ and 400, respectively. }
	\label{fig:beta5}
\end{figure}    
\end{document}